\documentclass[twocolumn,twocolappendix]{aastex63}

\usepackage{amsmath,mathtools,mathrsfs,amssymb}
\usepackage{hyperref}
\usepackage{verbatim}
\usepackage{xspace}
\usepackage{graphicx}
\usepackage{longtable}
\usepackage[T1]{fontenc}

\newcommand{\ud}{\mathrm{d}}
\newcommand{\dd}[2]{\dfrac{\ud #1}{\ud #2}}
\newcommand{\hodgestar}{\raisebox{1pt}{$\star{}$}}
\newcommand{\bhspin}{a_*}
\newcommand{\rin}{{r_{\mathrm{in}}}}
\newcommand{\MSE}{\mathrm{MSE}}
\newcommand{\SSIM}{\mathrm{SSIM}}
\newcommand{\DSSIM}{\mathrm{DSSIM}}

\newcommand{\patoka}[0]{{\tt{PATOKA}}\xspace}
\newcommand{\iharm}[0]{{\tt{iharm}}\xspace}
\newcommand{\ibothros}[0]{{\tt{ibothros}}\xspace}
\newcommand{\ipole}[0]{{\tt{ipole}}\xspace}
\newcommand{\igrmonty}[0]{{\tt{igrmonty}}\xspace}
\newcommand{\symphony}[0]{{\tt{symphony}}\xspace}

\shorttitle{Illinois EHT Library}
\shortauthors{Wong et al.}

\defcitealias{EHTC_2019_1}{EHTC~I}
\defcitealias{EHTC_2019_2}{EHTC~II}
\defcitealias{EHTC_2019_3}{EHTC~III}
\defcitealias{EHTC_2019_4}{EHTC~IV}
\defcitealias{EHTC_2019_5}{EHTC~V}
\defcitealias{EHTC_2019_6}{EHTC~VI}
\defcitealias{EHTC_2021_7}{EHTC~VII}
\defcitealias{EHTC_2021_8}{EHTC~VIII}

\begin{document}

\title{{\tt{}PATOKA}: Simulating Electromagnetic Observables of Black Hole Accretion}
\shorttitle{PATOKA}

\correspondingauthor{George~N.~Wong}
\email{gnwong@ias.edu}

\author[0000-0001-6952-2147]{George~N.~Wong}
\affiliation{Department of Physics, University of Illinois, 1110 West Green Street, Urbana, IL 61801, USA}
\affiliation{School of Natural Sciences, Institute for Advanced Study, 1 Einstein Drive, Princeton, NJ 08540, USA}
\affiliation{Princeton Gravity Initiative, Princeton University, Princeton, New Jersey 08544, USA}
\affiliation{Illinois Center for Advanced Study of the Universe, 1110 West Green Street, Urbana, IL 61801, USA}

\author[0000-0002-0393-7734]{Ben~S.~Prather}
\affiliation{Department of Physics, University of Illinois, 1110 West Green Street, Urbana, IL 61801, USA}
\affiliation{Illinois Center for Advanced Study of the Universe, 1110 West Green Street, Urbana, IL 61801, USA}

\author[0000-0001-6765-877X]{Vedant~Dhruv}
\affiliation{Department of Physics, University of Illinois, 1110 West Green Street, Urbana, IL 61801, USA}
\affiliation{Illinois Center for Advanced Study of the Universe, 1110 West Green Street, Urbana, IL 61801, USA}

\author[0000-0001-8939-4461]{Benjamin~R.~Ryan}
\affil{CCS-2, Los Alamos National Laboratory, P.O.~Box 1663, Los Alamos, NM 87545, USA}

\author[0000-0002-4661-6332]{Monika~Mo\'scibrodzka}
\affil{Department of Astrophysics/IMAPP, Radboud University, P.O. Box 9010, 6500 GL
Nijmegen, The Netherlands}

\author[0000-0001-6337-6126]{Chi-kwan~Chan}
\affil{Steward Observatory and Department of Astronomy, University of Arizona, 933 N. Cherry Ave., Tucson, AZ 85721, USA}
\affil{Data Science Institute, University of Arizona, 1230 N. Cherry Ave., Tucson, AZ 85721, USA}

\author[0000-0002-2514-5965]{Abhishek~V.~Joshi}
\affil{Department of Physics, University of Illinois, 1110 West Green Street, Urbana, IL 61801, USA}
\affil{Illinois Center for Advanced Study of the Universe, 1110 West Green Street, Urbana, IL 61801, USA}

\author[0000-0003-0381-1039]{Ricardo~Yarza}
\affil{Department of Astronomy and Astrophysics, University of California, Santa Cruz, CA 95064, USA}

\author[0000-0001-5287-0452]{Angelo~Ricarte}
\affil{Center for Astrophysics | Harvard \& Smithsonian, 60 Garden Street, Cambridge, MA 02138, USA}
\affil{Black Hole Initiative, 20 Garden Street, Cambridge, MA 02138, USA}

\author[0000-0002-8847-5275]{Hotaka~Shiokawa}
\affil{Rakuten Institute of Technology, 2 South Station, Suite 400, Boston, MA 02110, USA}

\author[0000-0003-4353-8751]{Joshua~C.~Dolence}
\affil{CCS-2, Los Alamos National Laboratory, P.O.~Box 1663, Los Alamos, NM 87545, USA}

\author[0000-0003-3547-8306]{Scott C. Noble}
\affiliation{Gravitational Astrophysics Lab, NASA Goddard Space Flight Center, Greenbelt, MD 20771, USA}

\author[0000-0002-9222-599X]{Jonathan~C.~McKinney}
\affil{H2O.ai, 2307 Leghorn Street, Mountain View, CA 94043, USA}

\author[0000-0002-0393-7734]{Charles~F.~Gammie}
\affiliation{Department of Physics, University of Illinois, 1110 West Green Street, Urbana, IL 61801, USA}
\affiliation{Illinois Center for Advanced Study of the Universe, 1110 West Green Street, Urbana, IL 61801, USA}
\affiliation{Department of Astronomy, University of Illinois, 1002 West Green Street, Urbana, IL 61801, USA}
\affiliation{National Center for Supercomputing Applications, 605 East Springfield Avenue, Champaign, IL 61820, USA}

\begin{abstract}

The Event Horizon Telescope (EHT) has released analyses of reconstructed images of horizon-scale millimeter emission near the supermassive black hole at the center of the M87 galaxy. Parts of the analyses made use of a large library of synthetic black hole images and spectra, which were produced using numerical general relativistic magnetohydrodynamics fluid simulations and polarized ray tracing.
In this article, we describe the \patoka pipeline, which was used to generate the Illinois contribution to the EHT simulation library. We begin by describing the relevant accretion systems and radiative processes. We then describe the details of the three numerical codes we use, \iharm, \ipole, and \igrmonty, paying particular attention to differences between the current generation of the codes and the originally published versions. Finally, we provide a brief overview of simulated data as produced by \patoka and conclude with a discussion of limitations and future directions.

\end{abstract}


\section{Introduction}\label{sec:intro}

\begin{figure*}
\centering
\includegraphics[width=\linewidth]{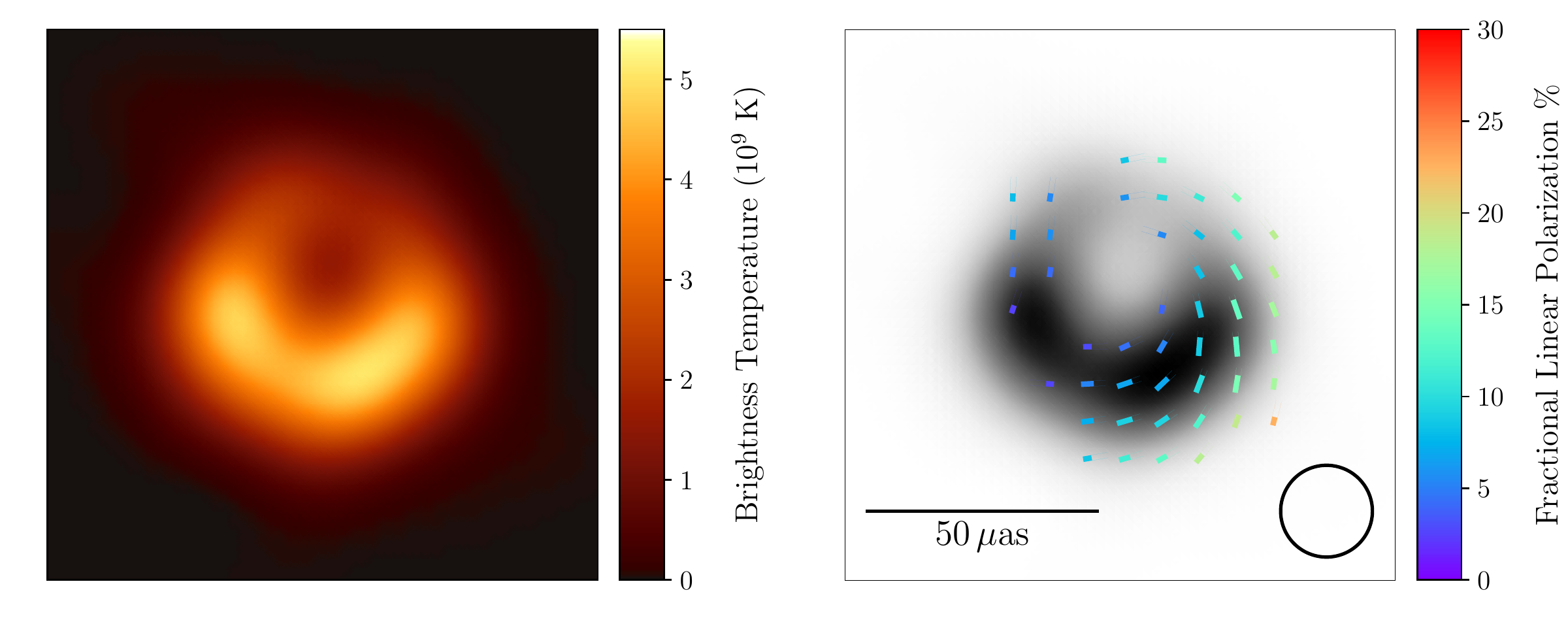}
\caption{
Image of the supermassive black hole at the center of the Messier 87 galaxy at $230\,$GHz produced by the Event Horizon Telescope Collaboration from observations taken on April 11, 2017. Left: total intensity of light shown in units of brightness temperature $T_b = S \lambda^2 / 2 k \Omega$, where $S$ is the (total) flux density, $\lambda$ is the observing wavelength, $k$ is the Boltzmann constant, and $\Omega$ is the solid angle of the observing element. Right: linear polarimetric data from the same observation plotted atop gray-scale total intensity as in the left panel. The orientation of each tick corresponds to the electric-vector position angle projected onto the image plane and its color encodes the local fractional linear polarization. Tick marks are only shown in regions where the local total intensity is $> 1/10$ its maximum value and fractional linear polarization is $> 1/5$ its maximum. The FWHM of the largest Gaussian blurring kernel used in image reconstruction is represented as a circle with diameter $20\,\mu$as in the bottom right of the panel. See \citetalias{EHTC_2019_4} and \citetalias{EHTC_2021_7} for more details.
}
\label{fig:eht_splash}
\end{figure*}

The Event Horizon Telescope (EHT) is a globe-spanning network of millimeter wavelength observatories that is capable of imaging two supermassive black holes at event-horizon-scale resolutions. In 2019, the EHT published the first total intensity images of horizon-scale emission from the center of the giant elliptical galaxy Messier~87
\citep[][hereafter EHTC~I--VI]{EHTC_2019_1,EHTC_2019_2,EHTC_2019_3,EHTC_2019_4,EHTC_2019_5,EHTC_2019_6}, and in 2021, the first linearly polarized images of the same source \citep[][hereafter EHTC~VII--VIII]{EHTC_2021_7,EHTC_2021_8}. Representative images from the publications are reproduced in Figure~\ref{fig:eht_splash}.
Improvements to the EHT including new antennas, increased sensitivities, and support for measurements at different frequencies lie on the horizon.

The EHT operates at millimeter wavelengths, and the first images of the presumed black hole at the center of Messier 87 (hereafter M87) were produced at $230\,$GHz ($1.3$ mm; the EHT observes at two $2\,$GHz-wide bands centered at $227.1$ and $229.1\,$GHz, \citetalias{EHTC_2019_1}).
The reconstructed images exhibit an asymmetric ring-like structure with diameter $42\, \pm \,3 \, \mu$as, which is consistent with the image of the shadow of a black hole with mass $\left(6.5 \,\pm\, 0.7\right) \, \times 10^9 M_\odot$ in the background of emission produced by hot, accreting plasma, as predicted by general relativity\footnote{Assuming a distance of $16.8$ Mpc (see \citetalias{EHTC_2019_1} for more detail).} \citepalias{EHTC_2019_1,EHTC_2019_5,EHTC_2019_6}. EHT modeling connected the ring to the effects of strong lensing of near-horizon emission and linked the orientation of the brightness asymmetry to the motion of the emitting plasma and the spin of the hole (\citetalias{EHTC_2019_5}; see also \citealt{wong_2021_JetDiskBoundaryLayer}). 
The data also showed a peak in linear polarization along the southwestern segment of the ring that is consistent with the presence of organized, poloidal magnetic fields in the emission region. The overall low fractional linear polarization suggests scrambling of the polarimetric signal on small scales; this scrambling is attributed to internal Faraday rotation (\citetalias{EHTC_2021_8}; see also \citealt{moscibrodzka_2017_FaradayRotationGRMHD,jimenezrosales_2018_ImpactFaradayEffects,ricarte_2020_DecomposingInternalFaraday}). Combined with data from other sources, the aggregate EHT data were used to infer constraints on the spin, mass, and accretion rate, as well as the magnetic-field strength and structure in the M87 accretion system \citepalias{EHTC_2019_1,EHTC_2021_8}. The theory constraints rely in part on understanding the complicated relationship between the detailed image features and the underlying spacetime geometry and accretion flow.

The EHT analysis relies in large part on a library of synthetic observations. This library comprises a set of numerical simulations of magnetized, relativistic black hole accretion flow models and accompanying polarized ray-traced simulated images and spectral energy distributions. These outputs can then be processed into synthetic observations via pipelines that simulate the effects of the observing process, e.g., {\tt{eht-imaging}} \citep{chael_2016_HighresolutionLinearPolarimetric}, {\tt{SYMBA}} \citep{roelofs_2020_SYMBAEndtoendVLBI}, or THEMIS \citep{broderick_2020_themis}.
Synthetic observations serve two primary functions: they can be used to validate image reconstruction procedures, and they can be used in the forward modeling pipelines that compare the observational and synthetic data to infer physical parameters of the system.

\begin{figure*}
\centering
\includegraphics[width=0.95\textwidth]{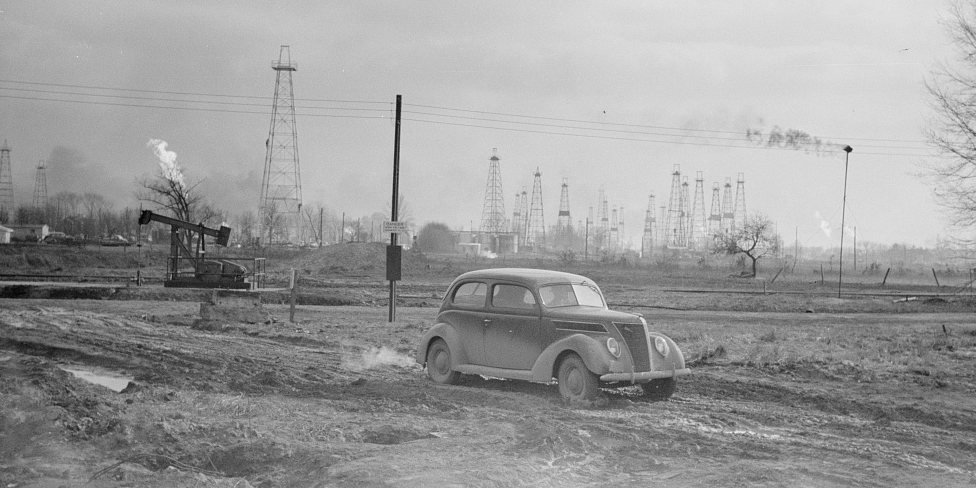}
\caption{Photograph of an oil field near Patoka in Marion County, Illinois published in February 1940. The \patoka pipeline is named for the Patoka Oil Terminal in Patoka, Illinois, which connects several oil pipelines in the second Petroleum Administration for Defense District. Photo credit: Arthur Rothstein/Farm Security Administration, LC-DIG-fsa-8a13148.}
\label{fig:patoka}
\end{figure*}

General relativistic magnetohydrodynamics (GRMHD) simulations are a mainstay in analysis of black hole accretion flows (e.g., \citealt{gammie_2003_harm,delzanna_2007_ECHOEulerianConservative,mignone_2007_PLUTONumericalCode,narayan_2012_sane,sadowski_2013_koral,white_2016_ExtensionAthenaCode,porth_2017_BHAC,liska_2018_FormationPrecessingJets}), and many general relativistic ray-tracing codes have been developed for both imaging \citep[e.g.,][building on earlier work by \citealt{mihalas_1984_FoundationsRadiationHydrodynamics}]{zane_1996_GeneralRelativisticRadiative,fuerst_2004_RadiationTransferEmission,noble_2007_ibothros,psaltis_2012_RaytracingAlgorithmSpinning,younsi_2012_GeneralRelativisticRadiative,chan_2013_gray,dexter_2016_grtrans,chan_2018_gray2,moscibrodzka_2018_ipole,pihajoki_2018_GeneralPurposeRay,kawashima_2021_RAIKOUGeneralRelativistic} and the generation of spectra \citep[e.g.,][]{dolence_2009_grmonty,moscibrodzka_2009_RadiativeModelsSGR,dolence_2012_NearinfraredXRayQuasiperiodic,schnittman_2013_montecarlo,zhang_2019_corona_mc,moscibrodzka_2020_polmonty,kawashima_2021_RAIKOUGeneralRelativistic}.

In this article, we describe the \patoka modeling pipeline, which was used for the Illinois contribution of fluid accretion models, images, and spectra to the EHT library. The details of this pipeline are messy, like the details of the pipelines used for hydrocarbon extraction around downstate Illinois's Patoka oil terminal (see Figure~\ref{fig:patoka}), for which our pipeline is named. In this paper, we describe some of the messy details, including differences between the published code descriptions and the versions that were used in generating the library. 
In Section~\ref{sec:theorybackground}, we provide a summary of the theoretical underpinnings of the accretion model and radiation physics. In Section~\ref{sec:codereview}, we review the details of the numerical codes used to perform the simulations and ray tracing. In Section~\ref{sec:results}, we present some results that were generated with the pipeline. We end with a discussion of future directions in Section~\ref{sec:discussion}.

\section{Theoretical Background}
\label{sec:theorybackground}

We begin with an overview of the theoretical considerations associated with library generation. We cover the parameters of the black hole accretion model and review the equations governing radiative transfer in hot astrophysical plasmas.

\subsection{Black Hole Accretion}

The supermassive black holes at the centers of Messier 87 and the Galactic Center (Sgr~A*) are associated with compact radio sources
\citep[see, e.g.,][]{lyndenbell_1969_GalacticNucleiCollapsed,balick_1974_IntenseSubarcsecondStructure,doeleman_2008_sgra,fish_2011_sgra} and are the primary targets for the EHT. M87 and Sgr~A* are usually described as low-luminosity active galactic nuclei (see \citealt{greene_2007_NewSampleLowMass,yuan_2014_HotAccretionFlows} for reviews), so the accretion onto their central black holes is expected to proceed at a low rate $\dot{m} \equiv \dot{M} / \dot{M}_{\mathrm{Edd}} < 10^{-3}$ and therefore be radiatively inefficient (RIAF for radiatively inefficient accretion flow or ADAF for advection dominated accretion flow; see \citealt{ichimaru_1977_BimodalBehaviorAccretion,narayan_1994_AdvectiondominatedAccretionSelfsimilar,narayan_1995_ADAF,quataert_1999_SpectralModelsAdvectiondominated,yuan_2003_RIAFSgra}). We express the mass accretion rate $\dot{m}$ in terms of the Eddington accretion rate $\dot{M}_{\mathrm{Edd}} \equiv L_{\mathrm{Edd}} / \left( \eta c^2\right)$, where $c$ is the speed of light, the nominal accretion efficiency is chosen to be $\eta = 0.1$, and the Eddington luminosity is $L_{\mathrm{Edd}} =  4 \pi G M m_p c / \sigma_{T}$, where $G$ is Newton's gravitational constant, the mass of a proton is $m_p$, the Thomson cross section is $\sigma_T$, and the mass of the central black hole is $M$.

For orderly accretion to occur, the gravitational binding energy of the system must be dissipated as matter falls from large to small radii. In RIAF models, the gravitational binding energy of the flow is converted into heat and either advected across the event horizon or lost in mechanical outflows. RIAFs are typically modeled as geometrically thick disks of relativistically hot, magnetized plasma \citep{rees_1982_IonsupportedToriOrigin,narayan_1995_ADAF,narayan_1995_ExplainingSpectrumSagittarius,reynolds_1996_QuiescentBlackHole,yuan_2002_NGC4258Jetdominated,dimatteo_2003_AccretionSupermassiveBlack,yuan_2014_HotAccretionFlows}. M87 also supports a relativistic jet that extends to kiloparsec scales and has an estimated jet power of $\approx 10^{42} - 10^{45}$ erg/s \citep{stawarz_2006_m87jet_hst1,degasperin_2012_m87jetpower,prieto_2016_CentralParsecsM87}. Whether or not Sgr~A* also launches a jet is less certain \citep[e.g.,][]{falcke_2000_JetModelSgr,yusefzadeh_2006_FlaringActivitySagittarius,markoff_2007_HowHideLargescale,falcke_2009_JetlagSagittariusWhat,brinkerink_2015_ALMAVLAMeasurements,issaoun_2019_SizeShapeScattering,brinkerink_2021_PersistentTimeLags}.

The Coulomb mean free paths of the ions and electrons in typical RIAF models are much larger than the typical length scales in the disks \citep[see][]{mahadevan_1997_AreParticlesAdvectiondominated}, so the accreting plasma should be collisionless, suggesting that nonideal processes may be important. Particle-in-cell (PIC) simulations \citep[e.g.,][]{kunz_2014_FirehoseMirrorInstabilities,riquelme_2015_ParticleincellSimulationsContinuously,sironi_2015_ElectronHeatingIon}, measurements of the solar wind, and phenomena like collisionless shocks infer that instabilities may increase the effective particle collision rate, allowing nonideal effects to be treated as a small correction (see \citealt{foucart_2016_EvolutionAccretionDiscs} and references therein). Extended GRMHD simulations that include these small corrections have shown that the low collisionality does not significantly alter the time-averaged structure of the flow \citep[e.g.,][]{foucart_2017_HowImportantNonideal}.

Kinetic plasma physics phenomena may ultimately drive the system out of equilibrium in ways that our methods do not account for; however, global general relativistic kinetic simulations of ion--electron plasmas in black hole accretion flows at realistic accretion rates are prohibitively expensive and have never been run. Either way, differences between electron and ion heating mechanisms and the absence of an efficient coupling mechanism between the two species implies at least that the electrons and ions in the flows are likely described by different temperatures (see \citealt{shapiro_1976_twotemp,mahadevan_1997_AreParticlesAdvectiondominated,quataert_1998_ParticleHeatingAlfvenic,ressler_2015_ElectronThermodynamicsGRMHD,zhdankin_2021_ProductionPersistenceExtreme}).

Analytic models are useful in understanding the broad characteristics of accretion flows (e.g., \citealt{novikov_1973_diskmodel,shakura_1973_diskmodel,abramowicz_1978_RelativisticAccretingDisks,kozlowski_1978_AnalyticTheoryFluid,narayan_1994_AdvectiondominatedAccretionSelfsimilar,yuan_2003_RIAFSgra,broderick_2006_FrequencydependentShiftImage,komissarov_2006_MagnetizedToriKerr,broderick_2011_EvidenceLowBlack,abolmasov_2018_hotthinaccretiondiscs}), but they provide an incomplete picture of the dynamics. For example, the analytic treatments often propose or rely on thin-disk models, do not explicitly include magnetic fields, and do not account for time dependence in the flows; yet the development of turbulence and shocks make an analytic treatment particularly challenging.\footnote{The magnetorotational instability of \citet[][MRI]{balbus_1991_mri} may play a crucial role in facilitating the angular momentum transport necessary for accretion in steady disks, whose horizon-scale flows are not choked, cf.~MAD flows in \S{}\ref{sec:magflux}. Even in MAD flows, the MRI may play a role in transient configurations when weaker magnetic fields thread the bulk accretion flow.}

Since a large part of the millimeter emission produced in a RIAF flow is likely produced near the horizon (e.g., \citealt{moscibrodzka_2009_RadiativeModelsSGR}), a full general relativistic treatment of the system is necessary to adequately recover the precise features of the model.
GRMHD simulations serve as a practical method for studying the plasma dynamics in detail. GRMHD simulations evolve the distribution of plasma and electromagnetic energy through time and produce a time-series description of the plasma state. Global GRMHD simulations have been widely used to study RIAF accretion and have been shown to reproduce the jet powers of the \citet{blandford_1977_bz} model (e.g.,  \citealt{koide_1999_RelativisticJetFormation,devilliers_2003_grmhdtori,gammie_2003_harm,mckinney_2004_harmbz,narayan_2012_sane}).
Numerical accretion models have been shown to be consistent with various observational constraints provided by jet powers, variability, and the sub-mm spectrum (e.g., \citealt{noble_2007_ibothros,moscibrodzka_2009_RadiativeModelsSGR,dexter_2010_SubmillimeterBumpSgr,dexter_2012_SizeJetLaunching,shcherbakov_2012_SagittariusAccretionFlow,dexter_2013_TiltedBlackHole,moscibrodzka_2014_ObsRIAF,dexter_2020_ParameterSurveySgr}). The consistency between simulations and the recent horizon-scale images of M87 published by the EHT further supports the RIAF/ADAF thick-disk picture \citepalias{EHTC_2019_5,EHTC_2021_8}.

At low accretion rates, radiative cooling has a negligible impact on the fluid dynamics (\citealt{dibi_2012_radcooling,ryan_2017_RadiativeEfficiencySpectra} and see also \citealt{yoon_2020_SpectralImagingProperties}), so the evolution is invariant under rescalings of both the metric length $GM/c^2$ and the mass density. Ideal GRMHD simulations thus only introduce the following physical parameters: the angular momentum of the system and the magnetic flux near the event horizon. 

\subsubsection{Angular momenta and tilted disks}

The angular momentum of the central black hole $J$ is often expressed in terms of the dimensionless black hole spin parameter $\bhspin \equiv J c / G M^2$, with $\left|\bhspin\right| \le 1$. The angular momentum vectors of the black hole and the accretion flow may be misaligned; the angular separation between the two axes is called tilt.
Although there are plausible scenarios that produce accretion flows with zero tilt, there is at present no way of rejecting models with strong or even maximal (180 degrees, i.e., retrograde) tilt. In fact, retrograde accretion is not particularly unexpected; it exists and has been resolved in the context of stellar disks (see \citealt{young_2020_AtomicHydrogenClues} and references therein).

Systems with nonzero tilt may have clear observational signatures. \citet{fragile_2008_EpicyclicMotionsStanding,dexter_2013_TiltedBlackHole,shiokawa_2013_phdt,white_2019_TiltedDisksBlack,chatterjee_2020_ObservationalSignaturesDisc} found that tilted accretion flows produce a two-arm shock that would significantly alter the morphology of radio images and the shape of the spectrum. Some simulations suggest that the inner regions of tilted disks ultimately align with the central black hole via an analogue of the \citet{bardeen_1975_LenseThirringEffectAccretion} effect \citep{mckinney_2013_AlignmentMagnetizedAccretion,liska_2018_FormationPrecessingJets}.

In our galaxy, there is strong evidence that Sgr~A* was recently in an active period of higher accretion \citep{totani_2006_RIAFInterpretationHigher,ponti_2013_TracesActivityGalactic}, and other observational studies and simulations report evidence of clumpiness within a few parsecs of Sgr~A* in the circumnuclear disk \citep{monterocastano_2009_GasInfallSgr,blank_2016_InnerCavityCircumnuclear}. Simulated mass feeding in the Galactic Center via magnetized stellar winds by \citet{ressler_2018_HydrodynamicSimulationsInner,ressler_2020_SurprisinglySmallImpact,ressler_2020_InitioHorizonscaleSimulations} found that a wide variety of different initial configurations led to the development of similar nondisk-like accretion flows with strongly poloidal horizon-scale magnetic-field configurations. Evidently, the angular momentum of the boundary condition that supplies the inner accretion flow with mass is likely to be variable on galactic timescales \citep[e.g.,][]{cuadra_2006_GalacticCentreStellar}.

More generally, a nontrivial time-dependent boundary feeding condition is not unlikely, as there is a large discrepancy between the long timescale that governs changes to properties of the black hole and the shorter ones over which the local environment of the black hole (i.e., the environment that governs the  feeding) changes. Thus, the bulk angular momentum of the accreting matter (far from the horizon) may exhibit a time-dependent tilt with respect to the spin angular momentum of the central black hole.

Time-dependent tilts have been invoked to explain the wobble observed in some relativistic jets \citep[see][]{natarajan_1999_WarpedDiscsDirectional}.  These arguments find extra support in the analysis of episodic jet variations and spectral analyses of the broad Fe line \cite[e.g.,][]{hjellming_1995_EpisodicEjectionRelativistic,rout_2020_RetrogradeSpinBlack}. Theoretical studies that consider the initial spin distribution of supermassive black holes \citep{deluca_2019_InitialSpinProbability} and simulate black hole growth via accretion and mergers in hierarchical galaxy formation models also often favor random alignment between the black holes and their disks \citep[see, e.g.,][]{volonteri_2005_DistributionCosmicEvolution}.

The prograde/retrograde dichotomy has also been used to explain differences between radio-loud and radio-quiet active galactic nuclei sources (\citealt{garofalo_2009_SignaturesBlackHole,garofalo_2010_EvolutionRadioloudActive}, but see \citealt{tchekhovskoy_2012_ProgradeRetrogradeBlack}). Some observational studies have claimed detection of retrograde accretion systems: \citet{morningstar_2014_SpinBlackHole}, \citet{chen_2016_SpinBlackHole}, and \citet{mikhailov_2018_CriteriaRetrogradeRotation} presented criteria for classifying systems as retrograde and singled out several known systems based on emission models and low estimated Eddington ratios.

\subsubsection{Magnetic flux}
\label{sec:magflux}

In ideal MHD, accreting plasma carries magnetic-field lines with it as it falls onto a central black hole. If the magnetic polarity is constant over time, then field lines will accumulate and increase the magnetic pressure near the hole, eventually saturating when the magnetic pressure is large enough to counterbalance the inward ram pressure of the accretion flow. The amount of magnetic flux threading the event horizon thus qualitatively divides accretion flows into two categories: the magnetically arrested disk (MAD) state \citep{bisnovatyi-kogan_1974_AccretionMatterMAD,igumenshchev_2003_ThreedimensionalMagnetohydrodynamicSimulations,narayan_2003_MagneticallyArrestedDisk}, in which the magnetic pressure is large and the large-scale component of the magnetic field is dynamically important; and the alternate standard and normal evolution (SANE) state \citep{narayan_2012_sane,sadowski_2013_EnergyMomentumMass}.

The magnetic flux threading one hemisphere of the black hole horizon $\int \ud A \cdot \mathbf{B}$ is
\begin{align}
    \int\limits_{\substack{r = r_{\mathrm{eh}} \\ \mathrm{hemisphere}}} \!\!\!\!\!\!\!\! \hodgestar F^{rt} \sqrt{-g} \, \ud \theta \ud \phi, \label{eqn:phibh}
\end{align}
where $g$ is the determinant of the covariant metric and the Hodge dual of the electromagnetic Faraday field tensor $F^{\alpha\beta}$ is
\begin{align}
    \hodgestar F^{\mu\nu} = \dfrac{1}{2} \epsilon^{\mu\nu\alpha\beta} F_{\alpha\beta}.
\end{align}
Here, we have used the Levi--Civita tensor $\epsilon^{\mu\nu\alpha\beta} = - \left[\mu\nu\alpha\beta\right] / \sqrt{-g}$, which differs from the permutation symbol $\left[\mu\nu\alpha\beta\right] \in \left\{0, +1, -1\right\}$ by a factor of $\sqrt{-g}$ and sign.\footnote{We let Greek indices run over all four dimensions $(0,1,2,3)$ and use Roman indices for the three spatial ones $(1,2,3)$.} We compute a proxy for the magnetic flux as
\begin{align}
    \Phi_{\mathrm{BH}} = \ \dfrac{1}{2} \!\!\! \int\limits_{\;\;r = r_{\mathrm{eh}}} \!\!\!\! \left| \hodgestar F^{rt} \right| \sqrt{-g} \, \ud \theta \ud \phi . \label{eqn:phiapprox}
\end{align}
Equation~\ref{eqn:phiapprox} approximates Equation~\ref{eqn:phibh} when the field is well-organized on the horizon. The flux $\Phi_{\mathrm{BH}}$ is conventionally normalized by $\dot{M}$
\begin{align}
    \phi \equiv \dfrac{ \Phi_{\mathrm{BH}} }{ \sqrt{ \dot{M} r_g^2 c } },
\end{align}
where $r_g \equiv GM/c^2$ is the gravitational radius of the hole. In the rationalized Lorentz--Heaviside units used in GRMHD simulations, $\phi$ saturates at $\phi_c \approx 15$ (see \citealt{tchekhovskoy_2011_EfficientGenerationJets} and \citealt{porth_2019_grmhdcomp}).\footnote{In Lorentz--Heaviside units, the vacuum permittivity and permeability are $\epsilon_0 = \mu_0 = 1$ as in Gaussian units, but the strength of the electric and magnetic fields (and the reciprocal of the fundamental charge) differs from Gaussian units by a factor of $\sqrt{4\pi}$. In Gaussian units, $\phi_c \approx 50$.} It is not at present possible to calculate what value of $\phi$ will result from an arbitrary field configuration without evolving the flow.

MAD accretion is spatially and temporally variable and tends to proceed in isolated, thin plasma streams that begin far from the hole. MAD flows are often punctuated by violent magnetic bubble eruption events that release excess trapped magnetic flux. Although events where magnetic flux escapes from the black hole are not fully understood, they have been interpreted as a magnetically driven convection between the disk and the hole \citep{marshall_2018_AngularMomentumTransport}.

Simulations suggest that the MAD vs.~SANE dichotomy is observationally encoded in signatures of polarization, variability, and the details of the jet--disk connection \citep[e.g.,][]{gold_2017_ProbingMagneticField,palumbo_2020_DiscriminatingAccretionStates,wong_2021_JetDiskBoundaryLayer}. Analysis of observations of radio-loud active galaxies and their jets are consistent with the MAD picture \citep[e.g.,][]{zamaninasab_2014_DynamicallyImportantMagnetic}, and recent analysis of EHT data also suggests that M87 accretion is MAD \citepalias{EHTC_2019_5,EHTC_2021_8}.

\subsubsection{Electron temperature and radiative effects}

GRMHD simulations typically assume that the dynamics of the accreting plasma can be recovered by treating the flow as a single, thermal fluid; they therefore track only the internal energy (or temperature) of the bulk plasma. In the real world, however, the plasma typically has a long Coulomb mean free path, so the electron temperature $T_e$ will not necessarily be equal to the fluid temperature \citep{shapiro_1976_twotemp,rees_1982_IonsupportedToriOrigin,narayan_1995_ADAF}. Conventionally, the electron temperature is assigned after the fact according to a model that assigns $T_e$ as a function of the local fluid parameters, such as the internal energy of the fluid and the local ratio of gas pressure to magnetic pressure, plasma $\beta \equiv P_{\mathrm{gas}}/P_{\mathrm{mag}}$ (e.g., \citealt{goldston_2005_SynchrotronRadiationRadiatively,moscibrodzka_2009_RadiativeModelsSGR,shcherbakov_2012_SagittariusAccretionFlow,moscibrodzka_2013_CoupledJetdiskModel,moscibrodzka_2014_ObsRIAF,
moscibrodzka_2016_rhigh}---see \citealt{anantua_2020_ComparisonAGNGRMHD} for a comparison of several different models). We discuss this approximation in more detail in Section~\ref{sec:approxpath}.

GRMHD simulations that recover the turbulent dynamics rely on implicit large eddy simulations (ILES; see \citealt{boris_1990_LargeEddySimulation,boris_1992_ILES_NewInsights,miesch_2015_iles,grete_2017_phdt}), wherein it is assumed that numerical dissipation at the grid scale emulates turbulence and dissipation below the grid scale and thus can be used to infer electron (and ion) heating rates. \citet{ressler_2015_ElectronThermodynamicsGRMHD,sadowski_2017_RadiativeTwotemperatureSimulations} described a method to track numerical dissipation in relativistic magnetohydrodynamics simulations for this purpose. Alternative descriptions of (nonideal) fluid mechanics have been proposed with direct treatments of viscosity and resistivity; such methods explicitly capture the dissipation.

In the systems we consider, plasma is typically energized through a combination of collisionless shocks, magnetic reconnection, and relativistic turbulence. For shocks, interactions between the charged particles and the magnetic field set the length scale (i.e., via the gyroradius), and the strongly nonequilibrium perturbations produce irreversible heating of the electrons, e.g., via first-order Fermi acceleration. In regions where reconnection occurs, the electrodynamics within the resultant current sheets do work on the plasma and transform magnetic energy into heat. Finally, turbulent cascades carry energy from large scales to small ones, ultimately energizing particles in the plasma at the smallest of scales. The energization mechanisms can produce anisotropic and highly nonthermal particle energy distributions that may differ between the different components of the plasma. These different energization mechanisms have been studied in different regimes and for different plasma compositions (e.g., \citealt{numata_2015_IonElectronHeating,sironi_2015_ElectronHeatingIon,sironi_2015_ElectronHeatingIona,boldyrev_2017_MagnetohydrodynamicTurbulenceMediated,loureiro_2017_CollisionlessReconnectionMagnetohydrodynamic,mallet_2017_DisruptionAlfvenicTurbulence,shay_2018_TurbulentHeatingDue,comisso_2018_ParticleAccelerationRelativistic,comisso_2021_pitchangle,nattila_2021_magheatturb,zhdankin_2021_ProductionPersistenceExtreme,ripperda_2022_plasmoid}).

Our pipeline is ultimately agnostic to the details of the energization mechanism. Given a model for how the dissipated energy is partitioned into electrons and ions (e.g., \citealt{sharma_2007_ElectronHeatingHot,howes_2010_PrescriptionTurbulentHeating,werner_2016_ExtentPowerlawEnergy,rowan_2017_ElectronProtonHeating,werner_2018_NonthermalParticleAcceleration,zhdankin_2018_NumericalInvestigationKinetic,kawazura_2019_ThermalDisequilibrationIons,rowan_2019_ElectronProtonHeating,kawazura_2020_IonElectronHeating,kawazura_2021_EnergyPartitionAlfvenic,zhdankin_2019_ElectronIonEnergization,kawazura_2021_EnergyPartitionAlfvenic}) and an assumption about the particle distribution functions (e.g., the electrons are well described by a thermal distribution), one can independently track and evolve the electron temperature. 

At low accretion rates $\dot{m}$, radiative losses are too slow to materially alter the internal energy of the flow (see, e.g., \citealt{sharma_2007_ElectronHeatingHot,dibi_2012_radcooling}) and nonradiative GRMHD simulations sufficiently recover the dynamics. As $\dot{m}$ approaches $10^{-5}$, however, synchrotron emission and Compton upscattering become important to the flow dynamics, and it is necessary to use a radiation-GRMHD scheme like {\tt KORAL} (\citealt{sadowski_2013_koral}, implementing a gray M1-closure scheme for the radiation), {\tt ebhlight} (\citealt{ryan_2015_bhlight,ryan_2019_ebhlight_software}, using a full Monte Carlo treatment of the radiation field), or {\tt MOCMC} (\citealt{ryan_2020_MOCMC}, using an adaptively refined Monte Carlo treatment of the radiation field). We note that M1-closure schemes may be inadequate where the radiation field is not symmetric with respect to the mean flux, for example in regions where there are multiple, distinct radiation sources. The transition regime in which radiation becomes important has been studied by, e.g., \citet{fragile_2009_grmhdwithrad,wu_2016_HotAccretionFlow,sadowski_2017_radtrans,sadowski_2017_RadiativeTwotemperatureSimulations,ryan_2017_RadiativeEfficiencySpectra}.

\subsection{Radiative Transfer}
\label{sec:radxfer}

Simulated $230\,$GHz images of supermassive black holes at sufficiently low accretion rates are often dominated by a ring-like feature with high brightness temperature $T_b$. In terms of observational parameters, $T_b = S \lambda^2 / 2 k \Omega$, where $S$ is an observed flux density describing power received per unit area per unit frequency, $\lambda$ is the observing wavelength, $k$ is the Boltzmann constant, and $\Omega$ is the solid angle of the observing element. In physical terms, $T_b$ is the temperature of a blackbody that would reproduce an observed intensity at a given frequency. The location of the ring is broadly consistent with the critical curve boundary that separates the null geodesics that terminate on the black hole event horizon from those that do not. 

Geodesics close to but outside of the critical curve are also lensed by the hole, and the set of all geodesics that are lensed enough to complete $n$ half-orbits around the hole defines the $n^{\mathrm{th}}$ photon subring (\citealt{johnson_2020_UniversalInterferometricSignatures}, see also \citealt{ohanian_1987_BlackHoleGravitational}); as $n \to \infty$, the subrings approach the critical curve. This half-orbit definition is nearly equivalent to a definition whereby subrings are indexed by the number of times they pass through the plane perpendicular to the symmetry axis of the spacetime, i.e., the disk midplane in aligned accretion flows, though differences may arise depending on whether the orbit is defined with respect to $\theta$, $\phi$, or $z$. Subrings are geometric regions on a black hole image and are therefore, like the black hole shadow and critical curve, formally unrelated to the visual appearance of the image, which depends on the properties of the underlying emission. Each subring corresponds to a relatively delayed, demagnified image of the universe. See Section~\ref{sec:em_obs} and Figure~\ref{fig:grrt_ringdecomp} in particular for a detailed example of the subring decomposition.

In the astrophysical scenarios we consider, emission drops off sharply with distance from the black hole, so images are dominated by emission produced near the hole. The emission near the black hole will thus produce a bright region in the direct $n=0$ image near the shadow. Since the image displayed by each next subring corresponds to a demagnified image of the previous one, the full image, which is produced by summing over each of the subring images, will display a sharp, logarithmically narrowing feature in intensity whose peak converges to the critical curve in the limit that the optical depth of the system is negligible. The feature seen in the full, composite image is known as the ``photon ring'' \citep{bardeen_1973_TimelikeNullGeodesics,luminet_1979_ImageSphericalBlack,johnson_2020_UniversalInterferometricSignatures,gralla_2020_LensingKerrBlack}, and the region it encircles is the black hole shadow. 

The location of the critical curve and the shape of the black hole shadow are controlled by the spacetime geometry. Strategies for constraining properties of the spacetime by measuring the critical curve via the shape of the photon ring or due to autocorrelations and subring image delays have been proposed (e.g., \citealt{falcke_2000_ViewingShadowBlack,takahashi_2004_ShapesPositionsBlack,bambi_2009_ApparentShapeSuperspinning,hioki_2009_MeasurementKerrSpin,amarilla_2010_NullGeodesicsShadow,amarilla_2013_ShadowKaluzaKleinRotating,tsukamoto_2014_ConstrainingSpinDeformation,younsi_2016_NewMethodShadow,mizuno_2018_CurrentAbilityTest,johnson_2020_UniversalInterferometricSignatures,medeiros_2020_ParametricModelShapes,olivares_2020_HowTellAccreting,wielgus_2020_ReflectionasymmetricWormholesTheir,wong_2021_BlackHoleGlimmer,hadar_2021_PhotonRingAutocorrelations,chesler_2021_LightEchosCoherent}).

Particle trajectories in curved spacetimes are determined by the geodesic equations
\begin{align}
    \dd{x^\alpha}{\lambda} &= k^\alpha \\
    \dd{k^\alpha}{\lambda} &= - {\Gamma^\lambda}_{\alpha\beta} k^\alpha k^\beta,
\end{align}
where $\Gamma$ is a Christoffel symbol, $\lambda$ the affine parameter, and $k^\alpha$ the photon wavevector. The frequency $\nu$ of the photon as measured in a frame with four-velocity $u^\alpha$ is in general given by $2 \pi \nu = - k^\alpha u_\alpha$, although the scale factor in this relation depends on the choice of normalization. In \patoka, we normalize the photon wavevector $k^\alpha$ such that the frequency is given by $\nu = \left( - k^\alpha u_\alpha \right) m_e c^2 / h$, where $h$ is Planck's constant.

The properties of the radiation field are often quantified in terms of the frequency- and angle-dependent \emph{specific intensity} of the light, $I_{\nu} =$ the amount of energy per area per solid angle per time per frequency.\footnote{The specific intensity is related to the brightness temperature by $I_{\nu} = 2 k T_b \nu^2 / c^2 = 2 \Theta_b m_e \nu^2$, where in the last term, we used the electron rest-mass energy to construct a dimensionless measure of brightness temperature $\Theta_b \equiv k T_b / \left(m_e c^2\right)$. Here we have taken the Rayleigh--Jeans limit with $h \nu \ll k T$, which is appropriate for mildly relativistic electrons and frequencies near $230\,$GHz.} Specific intensity changes along the geodesics according to interactions with the local matter in four ways: emission can increase the intensity of the light, scattering into the geodesic can increase the intensity of light, the intensity can decrease either due to absorption or scattering out of the geodesic (collectively called extinction), and the local plasma properties can mix linear and circular polarizations. The extinction term depends on the local intensity of the incoming light, the path-length along the geodesic $\ud s$, and an extinction coefficient, which itself depends on the local plasma parameters and, e.g., for synchrotron absorption, is a function of the angle the light ray makes with the magnetic field.

\subsubsection{Polarization}

In the supermassive black hole accretion systems targeted by the EHT, photons are emitted primarily by the synchrotron process (see, e.g., \citealt{yuan_2014_HotAccretionFlows}), when electrons in the accreting plasma are accelerated by magnetic fields. The local orientation of the magnetic field sets a preferred direction for the electromagnetic radiation, and thus synchrotron emission is, in general, polarized according to the direction of the magnetic field.

The full description of polarized light is often given in terms of the specific intensities of the Stokes parameters, which can be written in terms of the Stokes vector $\boldsymbol{S}_\nu \equiv \left(I_\nu, Q_\nu, U_\nu, V_\nu\right)$. Here, $I_\nu$ is the familiar total specific intensity of the light, $Q_\nu$ and $U_\nu$ represent the linearly polarized specific intensities, and $V_\nu$ represents the circularly polarized one. Not all light must be polarized, so $I_\nu^2 \ge Q_\nu^2 + U_\nu^2 + V_\nu^2$. The electric-vector position angle (EVPA),\footnote{We use the International Astronomical Union convention (see \citealt{hamaker_1996_UnderstandingRadioPolarimetry}) in which positive $Q$ is oriented north--south (vertically) and positive $U$ is oriented along the northeast--southwest direction (top left to bottom right). In this convention, EVPA is measured east of north, i.e., counterclockwise from vertical on the sky.} denoted $\chi$, is
\begin{align}
    \label{eqn:evpadefn}
    \chi \equiv \dfrac{1}{2} \arctan \dfrac{U}{Q}.
\end{align}
Physically, $\chi$ describes the angle the electric field oscillations make with respect to a fiducial direction. The orientation of the electric field exhibits a $\pi$-fold symmetry, so the representation of linear polarization $\chi$ is a pseudovector. The factor of one-half is due to this symmetry and can be seen in, e.g., the quadratic relationship between the Stokes parameters and the electric field vector $E$.

The polarized intensities are governed by the polarized radiative transfer equation. In order to produce polarized millimeter radio images, we need only treat the emission and absorption processes, since scattering contributes negligibly at the relevant frequencies and plasma parameters. Neglecting scattering, the polarized radiative transfer equation is
\begin{align}
    \label{eqn:polradxfer}
    \!\! \dd{}{s} \! \left( 
    \begin{matrix}
    I_\nu \\ Q_\nu \\ U_\nu \\ V_\nu
    \end{matrix}
    \right) = 
    \left(
    \begin{matrix}
    j_{\nu,I} \\ j_{\nu,Q} \\ j_{\nu,U} \\ j_{\nu,V}
    \end{matrix}
    \right) -
    \begingroup
    \setlength\arraycolsep{1pt}
    \begin{pmatrix}
    \alpha_{\nu,I} & \alpha_{\nu,Q} & \alpha_{\nu,U} & \alpha_{\nu,V} \\
    \alpha_{\nu,Q} & \alpha_{\nu,I} & \rho_{\nu,V} & - \rho_{\nu,U} \\
    \alpha_{\nu,U} & -\rho_{\nu,V} & \alpha_{\nu,I} & \rho_{\nu,Q} \\
    \alpha_{\nu,V} & \rho{\nu,U} & -\rho_{\nu,Q} & \alpha_{\nu,I}
    \end{pmatrix}
    \endgroup
    \left( 
    \begin{matrix}
    I_\nu \\ Q_\nu \\ U_\nu \\ V_\nu
    \end{matrix}
    \right),
\end{align}
where the emissivities $j_\nu$, absorptivities $\alpha_\nu$, and rotativities $\rho_\nu$ are frame-dependent quantities \citep{chandrasekhar_1960_RadiativeTransfer}. We work in a frame determined by the orientations of the wavevector and the magnetic field; this choice leads to $\alpha_U = \rho_U = 0$.

As polarized light travels through a magnetized plasma, several effects may cause $\chi$ to rotate. The first is due to parallel transport of the polarization vector in the curved geometry. The magnitude of this effect is a pure function of the underlying spacetime (see \citealt{pihajoki_2018_GeneralPurposeRay} and also see \citealt{gelles_2021_PolarizedImageEquatorial} for a quantitative study of the magnitude in the context of EHT-like sources). 
The others are due to interactions between the propagating light and the local plasma and are due to emission, absorption, and Faraday rotation and conversion. For example, in a magnetized plasma, the dielectric constant is a tensorial quantity, so components of the light with different polarizations propagate at different speeds. This magnetically induced birefringence produces a characteristic \emph{Faraday rotation} of $\chi$ that is a function of the plasma properties along the line of sight
\begin{align}
    \label{eqn:generalfaradayrotation}
    \Delta \chi &= \dfrac{1}{2} \int \ud s \, \rho_V \\ 
    &\approx \dfrac{1}{\nu^2} \dfrac{e^3}{2 \pi m_e^2 c^2} \int \ud s \, n_e  f_{\mathrm{rel}}(\Theta_e) B_{||},
\end{align}
where $\rho_V$ is the Faraday rotation coefficient.
In the second line, we have substituted a high-frequency expression for $\rho_V$ for the sub-to-mildly relativistic case; here, $n_e$ is the electron number density, $\Theta_e = k T_e / \left(m_e c^2\right)$ is the dimensionless electron temperature, $B_{||}$ is the component of the magnetic field along the line of sight, and
$f_{\mathrm{rel}} \approx \log \Theta_e / 2 \Theta_e$ for the $\Theta_e \gg 1$ case and $\approx 1$ otherwise (e.g., \citealt{rybicki_1979_RadiativeProcessesAstrophysics,quataert_2000_ConstrainingAccretionRate}, see also \citealt{shcherbakov_2008_PropagationEffectsMagnetized} for a discussion).
Evidently the observed polarization pattern is related to the structure of the magnetic field in a complicated way.

\citet{broderick_2004_semicovxfer} described a method to write the transport equation in terms of quantities that account for rotations of the observer frame along the line of sight. To provide a manifestly covariant description, \citet{gammie_2012_FormalismCovariantPolarized} produced an alternative formulation of the polarized transport equation in terms of a Hermitian coherency tensor $N^{\alpha\beta}$, which can be related to the Stokes parameters. The polarized transfer equation can then be written
\begin{align}
    k^\mu \, \nabla_\mu \, N^{\alpha\beta} = J^{\alpha\beta} + H^{\alpha\beta\gamma\delta} \, N_{\gamma\delta},
    \label{eqn:covariantpoltransport}
\end{align}
where $k^\mu \nabla_\mu$ is the derivative along the geodesic, $J^{\alpha\beta}$ is an emissivity tensor, and $H^{\alpha\beta\gamma\delta}$ is a tensor that accounts for absorption and generalized Faraday rotation.
Using this description to solve the polarized transport equation amounts to evaluating the nonrelativistic emissivities, absorptivities, and rotativities in the plasma frame, constructing $J^{\alpha\beta}$ and $H^{\alpha\beta\gamma\delta}$, and solving Equation~\ref{eqn:covariantpoltransport}. In particular, $J^{\alpha\beta}$ and $H^{\alpha\beta\gamma\delta}$ are constructed in the tetrad basis defined by the fluid four-velocity, the photon wavevector, and the orientation of the local magnetic field.

In some cases, we may solve an approximate form of the transfer equation that does not account for polarization. Beginning with Equation~\ref{eqn:polradxfer} and suppressing the polarized Stokes $Q,U,V$ terms produces
\begin{align}
    \dd{I_\nu}{s} &= j_{\nu,I} - \alpha_{\nu,I} I_\nu.
    \label{eqn:approxradxferflat}
\end{align}
This approximation is sometimes called \emph{unpolarized} transport, although it accounts for the total intensity of the light rather than only the unpolarized component. To recast the unpolarized transfer equation in a manifestly covariant form, notice that particle number $\ud N$ and phase space volume, $\ud^3 x \, \ud^3 p$, are conserved. Suppressing factors of $h$ and $c$, the phase space volume for a parcel of radiation at a frequency $\nu$ can be written $\ud^3 x \, \ud^3 p = \ud A \, \ud t \ \, \nu^2 \ud \nu \, \ud \Omega$. Thus, the quantity
\begin{align}
    \dfrac{I_\nu}{\nu^3} = \dfrac{1}{\nu^3} \dfrac{\nu \ud N}{\ud A \, \ud t \, \ud \nu \, \ud \Omega} = \dfrac{\ud N}{\ud A \, \ud t \, \nu^2 \ud \nu \, \ud \Omega} = \dfrac{\ud N}{\ud^3 x \, \ud^3 p}
\end{align}
is invariant. Similarly, $j_\nu / \nu^2$ and $\nu \alpha_\nu$ are invariant, so we can rewrite Equation~\ref{eqn:approxradxferflat} in covariant form as
\begin{align}
    \dd{}{s} \left(\dfrac{I_\nu}{\nu^3}\right) &= \left( \dfrac{j_{\nu,I}}{\nu^2} \right) - \left( \nu \, \alpha_{\nu,I} \right) \, \left( \dfrac{I_\nu}{\nu^3} \right).
    \label{eqn:approxradxfer}
\end{align}
We often write the invariant intensity $\mathcal{I}_\nu \equiv I_\nu / \nu^3$.

\subsubsection{Scattering}

Photons undergo Compton scattering as they travel through the plasma in the accretion flow. 
Compton scattering in the regime of interest here typically increases the energy of the scattered photon.
Collectively, these scattering events alter the spectrum; this is called \emph{Comptonization}. The Compton $y$ parameter gauges the significance of Comptonization:
\begin{align}
    y \equiv&\; N_{\mathrm{scatterings}} \times ({\scriptstyle{\mathrm{fractional\ energy\ gained/scattering}}}) \\ 
    \simeq &\; \tau_e \times  16 \Theta_e^2,
    \label{eq:comptonyrel}
\end{align}
where in the second line we have given an approximate form for the regime relevant to EHT sources. For a system of size $l$, the Thomson scattering optical depth is approximately
\begin{align}
    \tau_e \approx n_e \sigma_T l.
\end{align}

The rate of interaction between a photon with wavevector $k^\mu$ and massive particles in a distribution function $\ud n_m / \ud^3 p$ is
\begin{align}
    \dot{n}_{\mathrm{int}} &= \int \dfrac{\ud^3 p}{k^t} \, \dd{n_m}{^3 p} \left(- k_\mu p^\mu\right) \sigma_{m\gamma} \, c \\
    &= n_m \sigma_h \, c ,
\end{align}
where in the second line, we have used a ``hot cross section'' (see Appendix III of \citealt{canfield_1987_InverseComptonizationOnedimensional})
\begin{align}
    \sigma_h \equiv \int \ud^3 p \, \dd{n}{^3 p} \left( 1 - \mu_m \beta_m \right) \sigma_{m\gamma} \, c,
\end{align}
where we rewrite the dot product in terms of the particle speed in the plasma frame $\beta_m$ and the cosine of the angle between the particle momentum and the photon momentum $\mu_m$. For electron scattering, the cross section is the Klein--Nishina cross section, which is most conveniently written
\begin{align}
    \sigma_{\mathrm{KN}} = \sigma_T \dfrac{3}{4 \epsilon^2} \left( 2 + \dfrac{\epsilon^2 \left(1 + \epsilon\right)}{\left(1 + 2 \epsilon\right)^2} + \dfrac{\epsilon^2 - 2 \epsilon - 2}{2 \epsilon} \log\left(1 + 2 \epsilon\right) \right),
\end{align}
where the energy of the photon in the electron rest frame is $\epsilon = - p_\mu k^\mu / m_e$.

We use a Monte Carlo approach to process Compton scattering events. First, based on the incident photon, we use rejection sampling to draw an electron from the local electron distribution function (eDF), then we use rejection sampling again to sample the differential scattering cross section for that electron and incident photon in order to pick the energy and scattering angle that sets the wavevector for the scattered (outgoing) photon. We use the Thomson differential cross section for low-energy photons; otherwise we use the Klein--Nishina differential cross section.

\subsection{Connecting GRMHD and radiative transfer}

We produce the above-described electromagnetic observables (images and spectra) from the fluid simulations (which output a description of the fluid state) by performing radiative transfer calculations. In order to use the GRMHD output, the simulations must be scaled (assigned cgs units) and oriented, and an eDF must be specified.

\subsubsection{Scaling and orienting GRMHD output}

Unlike in GRMHD, the equations of radiative transfer are not scale invariant, so the numerical fluid data must be translated into physical units in order to perform the ray tracing. General relativistic radiative transfer (GRRT) introduces two scales: a length scale and a density scale.  The length scale $\mathscr{L}$ determines the absolute size of the accretion system and is often written in terms of the mass of the central black hole $\mathscr{L} = G M / c^2$.
The length scale sets the characteristic timescale, $\mathscr{T} = \mathscr{L}/c = GM/c^3$.

The density scale provides units to the fluid rest-mass density, internal energy, and magnetic-field strength while respecting the constraint that magnetization $\sigma = b^2 / \rho$ and plasma $\beta$ are independent of the choice of units. The density scale is specified as the ratio of a mass scale $\mathscr{M}$ to the volume scale determined by $\mathscr{L}$.\footnote{Note here that the mass scale is independent of the black hole mass because of scale separation---the accretion flow behaves as a test fluid and is effectively nonself-gravitating.
}
The density scale is chosen so that the simulated images match an observational constraint: the observed flux density must be correct. Since the simulated images have a limited field of view, it is possible for flows with large optical depths to produce diffuse, extended emission with image-integrated fluxes that are only consistent with the observation because of the limited image size (see Appendix~\ref{app:lmd_degeneracy}). The correct value of $\mathscr{M}$ is typically found via a root-finding procedure (see Appendix~\ref{app:lmd_degeneracy} for more detail). When possible, the resulting $\dot{M}$ is compared to the predictions of single-zone models (see \citetalias{EHTC_2019_5,EHTC_2021_8}) and observational estimates based on rotation measure (e.g., \citealt{bower_2003_InterferometricDetectionLinear,marrone_2006_SubmillimeterPolarizationSgr,marrone_2007_UnambiguousDetectionFaraday,kuo_2014_MeasuringMassAccretion}).

With these choices made, the state of the system is determined. Note that the choice of length and density scales is nearly degenerate with the choice of observer-to-source distance $\mathscr{D}$. We discuss this point in Appendix~\ref{app:lmd_degeneracy}.

Finally, the accretion flow must be oriented properly with respect to the camera location on Earth. This amounts to choosing two of the three Euler angles:
an inclination angle $i$ between the spin axis of the black hole and the elevation of the camera, and a position angle for the camera (the orientation that the projected spin axis makes on the image plane). Since we consider systems with zero tilt, our flows are statistically axisymmetric, so the third angle is negligible because there is no preferred azimuthal orientation of the disk about the spin axis of the system. The inclination and position angle are free parameters, although other observational information about the system, such as the orientation of a jet, may constrain both.

\subsubsection{Assigning electron temperatures}

Although our GRMHD simulations often make the approximation that the plasma is thermal and described by a single temperature, we account for the likely collisionless nature of the flow by allowing electron temperatures to deviate from the ion temperatures. In \patoka, when not using the entropy tracking procedure of \citet{ressler_2015_ElectronThermodynamicsGRMHD}, we assign electron temperatures according to the prescription of \citet[][see also \citealt{moscibrodzka_2017_FaradayRotationGRMHD}; \citetalias{EHTC_2019_5}]{moscibrodzka_2016_rhigh}, wherein the ion-to-electron temperature ratio is determined by the local plasma $\beta$. This prescription is motivated by the idea that electron heating may be relatively stronger in a strongly magnetized plasma (e.g., \citealt{quataert_1998_ParticleHeatingAlfvenic,quataert_1999_TurbulenceParticleHeating}), which can produce a sigmoidal dependence of the temperature ratio on $\beta$.

In \citet{moscibrodzka_2016_rhigh}, the temperature ratio is parameterized by $r_{\mathrm{low}}$, $r_{\mathrm{high}}$, and $\beta_{\mathrm{crit}}$, with
\begin{align}
    R \equiv T_i / T_e = r_{\mathrm{low}} \dfrac{1}{1 + \tilde{\beta}^2}  + r_{\mathrm{high}} \dfrac{\tilde{\beta}^2}{1 + \tilde{\beta}^2}
    \label{eqn:tptemodel},
\end{align}
where $\tilde{\beta} \equiv \beta/\beta_{\mathrm{crit}}$. We typically adopt $1 \lesssim r_{\mathrm{low}} \lesssim 10$ and $1 \lesssim r_\mathrm{high} \lesssim 160$ with $r_\mathrm{high} \geq r_\mathrm{low}$, based on more detailed studies of electron heating.

To recover the electron temperature from the total fluid internal energy, we partition the fluid internal energy into two components, associated with the electrons and the ions.\footnote{Some codes instead set the ion temperature equal to the fluid temperature, which overcounts the energy in the system.} Assuming an ideal gas equation of state, the energies associated with the ions and electrons will be
\begin{align}
    u_i &= (\hat{\gamma}_i - 1)^{-1} n_i k T_i \\
    u_e &= (\hat{\gamma}_e - 1)^{-1} n_e k T_e ,
\end{align}
where $\hat{\gamma}_i$ and $\hat{\gamma}_e$ are the adiabatic indices of the ions and electrons respectively. The ion and electron number densities are related to the total mass density by
\begin{align}
    n_e &= y \rho / m_p \\
    n_i &= z \rho / m_p,
\end{align}
where $y$ and $z$ are the number of electrons and nucleons per unionized atom, respectively.\footnote{Notice that $y$ and $z$ are not the conventional $Y$ and $Z$ mass fractions.} The electron temperature is thus
\begin{align}
    T_e &= \dfrac{m_p u \left(\hat{\gamma}_e - 1\right)\left( \hat{\gamma}_i - 1\right)}{k \rho \left(\left(\hat{\gamma}_i - 1\right) y + \left(\hat{\gamma}_e - 1\right) R z \right)} .
    \label{eqn:equationte}
\end{align}

The ions are typically nonrelativistic, so $\hat{\gamma}_i = 5/3$, and that the electrons are relativistic, so $\hat{\gamma}_e = 4/3$. Since we assign electron temperatures after the fact, we assume that $\hat{\gamma}_i, \hat{\gamma_e}$, and $\hat{\gamma}$ are constant across the simulation domain. This treatment is not entirely self-consistent, since the adiabatic index of the fluid should change depending on local contributions from the ion and electron fluid components (see \citealt{mignone_2007_eos_rmhd,choi_2010_MultidimensionalRelativisticHydrodynamic,mizuno_2013_RoleEquationState,shiokawa_2013_phdt,sadowski_2017_RadiativeTwotemperatureSimulations} for self-consistent treatments).

\subsubsection{Approximations and pathologies}
\label{sec:approxpath}

GRMHD schemes are not robust in regions with high magnetization $\sigma \gg 1$ or where plasma $\beta \ll 1$. In order to ensure numerical stability, $\sigma$ and $\beta$ are computed in each cell of the simulation domain for each time step, and mass or internal energy is injected into simulation zones to ensure that neither $\sigma$ nor $\beta^{-1}$ exceed some preset ceilings (see Section~\ref{sec:grmhd_models_and_floors} for more detail). We have typically varied the values of the ceilings across different trial-run simulations to ensure that they have minimal impact on the evolution of the flow. 

In black hole accretion simulations, the ceilings are generally triggered in the highly magnetized funnel regions near the poles, where we expect very low densities $\rho / \rho_{\mathrm{max}} \ll 1$. Although truncation errors dominate the evolution of the fluid internal energy in regions where the ceilings are activated, we expect that the plasma density in the jet will be low enough that it has negligible effect on the radiation. We therefore set the particle number density to zero
in regions with $\sigma > \sigma_{\mathrm{cut}}$. We typically set $\sigma_{\mathrm{cut}} = 1$. We also zero the particle number density within a few degrees of the poles, since some treatments of the polar boundary condition during the fluid evolution may cause $\sigma$ to artificially drop below unity there. The effect of the $\sigma$ cutoff is considered in \citet{chael_2019_TwotemperatureMagneticallyArrested}, \citetalias{EHTC_2021_8}, and Appendix~\ref{app:sigmacutoff}. In reality, the jet may be populated by an ``injected'' electron--positron pair plasma produced through photon--photon interactions in either pair cascades or pair drizzle (see Section~\ref{sec:rad_pairs}), although the plasma injection rate (and composition) will, in general, differ from the plasma injected due to floors.

\vspace{1em}

In images produced at edge-on inclinations, much of the emission travels through the disk on its way to the camera. At intermediate-to-large radii, the disk plasma has not had time to reach a turbulent quasi-equilibrium state whose average properties evolve only on the accretion timescale, so its state is largely a relic of the initial conditions. Thus, care must be taken that the polarization is not determined by Faraday rotation in the unequilibrated outer disk.  For the same reason, care must be taken that the polarization is not dependent on the size of the simulation domain (see, e.g., \citealt{ricarte_2020_DecomposingInternalFaraday}).

\vspace{1em}

Although we have assumed that the plasma is a pure electron--ion plasma, it is likely that electron--positron pairs populate parts of the domain. Pair plasmas produce different emission signatures \citep{anantua_2020_DeterminingCompositionRelativistic,emami_2021_PositronEffectsPolarized}, especially in polarization and in regions of the flow where the background plasma density is low. Furthermore, the presence of a pair plasma in low-density regions---like the jet---may influence the structure and dynamics of those regions.

\vspace{1em}

Finally, we use an ideal GRMHD scheme to evolve the accretion flow, which implicitly assumes that the plasma is in equilibrium and that the conductivity of the plasma is infinite, such that there are no unscreened electric fields. In fact, neither of these assumptions is likely to be valid globally across our simulation domain. It is unclear how strongly the ideal fluid assumption influences our results, but it is likely that the implicit assumption that the eDF is thermal and isotropic would have a significant effect on simulated observables. For example, kinetic simulations of pair plasmas infer strongly anisotropic, nonthermal eDFs \citep{comisso_2021_pitchangle,nattila_2021_magheatturb}, which can have significant consequences for polarimetric observables and even the distribution of emission across the domain, since emission is pitch-angle dependent.

There is yet limited consensus between different kinetic simulations, so it is not clear which model to use for subgrid phenomena and deviations from the ideal fluid assumption. We choose to address these uncertainties in part through our parametric treatment of the electron temperature assignment, which affords us great flexibility in translating simulation quantities into an eDF. Although our prescription is motivated, it cannot account for all of the potential deviations between our fluid treatment and a full kinetic treatment of the plasma distribution function. Understanding the differences between results from ideal (and extended) GRMHD fluid treatments and kinetic simulations of ion--electron--positron plasmas is important; however, it requires a detailed study that is beyond the scope of this paper.

\section{Code Detail}
\label{sec:codereview}

We now describe the \patoka pipeline, in which simulated observables are generated by ray tracing snapshots produced from GRMHD simulations of the black hole accretion flows. These two stages, GRMHD and GRRT, are separated for computational efficiency, since GRMHD simulations are costly and multiple radiation models can be applied to a single fluid snapshot without rerunning the fluid simulation. We describe the details of the three codes we use, placing an emphasis on differences between the versions we use and the code as described upon release. All codes used in \patoka compute metric derivatives via numerical differences and therefore nearly all coordinate dependence is encapsulated in a single line of code that specifies the line element.

\subsection{General relativistic magnetohydrodynamics}

We use the \iharm code (\citealt{gammie_2003_harm,noble_2006_PrimitiveVariableSolvers,noble_2009_nobleharm3d,prather_2021_iharm}) to integrate the equations of ideal GRMHD. \citet{porth_2019_grmhdcomp}
provides a comparison of contemporary GRMHD codes in the context of SANE accretion flows. \iharm is a conservative second-order explicit shock-capturing finite-volume code for arbitrary stationary spacetimes. \iharm is a descendant of {\tt{harm2d}}, and based on the {\tt{harm}} scheme of \citet{gammie_2003_harm}.

\iharm is designed to vectorize efficiently and achieves good performance and scaling on the systems used for this study. Validation and scaling tests are described in \citet{prather_2021_iharm} and show second-order convergence on a suite of test problems. The code is publicly available.\footnote{\url{https://github.com/afd-illinois/iharm3d}}

The governing equations of ideal GRMHD take the form of a hyperbolic system of conservation laws. In conservation form and written in a coordinate basis, the equations are
\begin{align}
\partial_t \left( \sqrt{-g} \rho u^t \right) &= -\partial_i \left( \sqrt{-g} \rho u^i \right), \label{eqn:massConservation}\\
    \partial_t \left( \sqrt{-g} {T^t}_{\nu} \right) &= - \partial_i \left( \sqrt{-g} {T^i}_{\nu} \right) + \sqrt{-g} {T^{\kappa}}_{\lambda} {\Gamma^{\lambda}}_{\nu\kappa},  \\
\partial_t \left( \sqrt{-g} B^i \right) &= - \partial_j \left[ \sqrt{-g} \left( b^j u^i - b^i u^j \right) \right], \label{eqn:fluxConservation}
\end{align}
with the constraint
\begin{align}
\partial_i \left( \sqrt{-g} B^i \right) &= 0, \label{eqn:monopoleConstraint} 
\end{align}
where the plasma is defined by its rest-mass density $\rho_0$, its four-velocity $u^\mu$, and $b^\mu$ is the magnetic field four-vector following (\citealt{mckinney_2004_harmbz}; see also \citealt{lichnerowicz_1967_RelativisticHydrodynamicsMagnetohydrodynamics}). Here, $\Gamma$ is a Christoffel symbol, and $i$ and $j$ denote spatial coordinates.
In Equations~\ref{eqn:fluxConservation} and~\ref{eqn:monopoleConstraint}, we use the ideal MHD condition\footnote{In ideal MHD, the plasma is assumed to have infinite conductivity so that the Lorentz force vanishes, with $\boldsymbol{E} + \boldsymbol{v} \times \boldsymbol{B} = 0$. In covariant form, this condition is $u_\mu F^{\mu\nu} = 0$.} $u_\mu F^{\mu\nu} = 0$ to express $F^{\mu\nu}$ in terms of $B^i \equiv \hodgestar F^{it}$ for notational simplicity. Notice that $B^i$ is the magnetic field as measured in the coordinate frame. The stress--energy tensor ${T^\mu}_\nu$ contains contributions from both the fluid and the electromagnetic field
\begin{align}
{T^{\mu}}_{\nu} &= \left( \rho + u + P + b^{\lambda}b_{\lambda}\right)u^{\mu}u_{\nu} + \left(P + \frac{b^{\lambda}b_{\lambda}}{2} \right){g^{\mu}}_{\nu} - b^{\mu}b_{\nu},
\label{eqn:mhdTensor}
\end{align}
where $u$ is the internal energy of the fluid and the fluid pressure $P$ is related to its internal energy through a constant adiabatic index $\hat{\gamma}$ with $P = \left(\hat{\gamma} - 1\right) u$. The four-velocity of the fluid is encoded in the three numbers $\tilde{u}^i$, whose definitions were chosen to improve numerical stability and are given in Appendix~\ref{app:harmprim}.

The equations are solved over a logically Cartesian, three-dimensional grid in arbitrary coordinates. Eight primitive variables,\footnote{Eight variables are saved for basic, ideal GRMHD. Extra variables may be tracked in extensions.} $\rho$, $u$, $\tilde{u}^i$, and $B^i$, are stored at the center of each grid zone and evolved forward in coordinate time (see Appendix~\ref{app:harmprim} for more detail). Fluxes are computed using the local Lax--Friedrichs method \citep{rusanov_1962_LLF}, and the divergence constraint of Equation~\ref{eqn:monopoleConstraint} is enforced using the Flux-CT scheme described in \citet[][see also \citealt{evans_1988_SimulationMagnetohydrodynamicFlows}]{toth_2000_ConstraintShockCapturingMagnetohydrodynamics}. The fifth-order WENO5 scheme \citep{jiang_1996_EfficientImplementationWeighted} is typically used for spatial reconstruction. The code is parallelized across contiguous domain chunks using MPI \citep{forum_1994_MPIMessagePassingInterface} and within each chunk using OpenMP \citep{dagum_1998_OpenMPIndustryStandardAPI}. \iharm imposes a constant floor on plasma $\beta$ and a ``geometric'' floor constraint on $\rho$ that varies with spatial coordinate; it imposes ceilings on $\sigma$, $\Theta_e$, the fluid velocity measured with respect to the coordinate frame, and optionally the logarithm of the fluid entropy $P_{\mathrm{gas}}/\rho^{\hat{\gamma}}$ (see Section~\ref{sec:grmhd_models_and_floors} for more detail). 

Simulations are typically run in augmented versions of the horizon-penetrating modified Kerr--Schild (MKS) coordinates introduced in \citet[][see Appendix~\ref{app:fmkscoords} for more detail]{gammie_2003_harm}. In MKS, the three spatial coordinates $x^1, x^2, x^3$ are direct functions of radius, latitude, and azimuth respectively. The inner radial boundary of the simulation is chosen to ensure that $\gtrsim 5$ zones lie within the event horizon. The outer edge of the boundary is typically chosen so that the torus lies comfortably within the simulation domain. We use outflow boundary conditions along the two radial boundaries, a periodic boundary condition in the azimuthal direction, and a pseudo-reflecting boundary condition at the two poles that mirrors the latitudinal $2$-components of the magnetic field and fluid velocity across the one-dimensional border.

\iharm supports the subgrid electron heating scheme of \citet{ressler_2015_ElectronThermodynamicsGRMHD} to track the local dissipation rate and then apportion some fraction of the dissipated energy to the electrons using any prescription that can be specified in terms of the fluid state. When the electron thermodynamics module is active, the code outputs two extra variables---functions of the total fluid entropy and the electron entropy---in addition to the principal eight primitive variables listed above.

\vspace{1em}

We initialize the fluid sector of our accretion disk simulations with a \citet[][hereafter FM]{fishbone_1976_RelativisticFluidDisks} torus. We perturb the initial conditions by sending $u \to u + \delta u$, with $\left| \delta u / u \right| \le u_{\rm jitter}$ drawn random uniformly per simulation zone. Typically $u_{\rm jitter} \approx 0.1$, although this value has been varied substantially between different simulations. This perturbation helps break the model symmetries and seed instabilities, including the magneto-rotational instability \citep{balbus_1991_mri}, and begin accretion onto the hole.\footnote{In 3D, a purely fluid torus is unstable to the \citet{papaloizou_1984_DynamicalStabilityDifferentially} instability.} The FM torus has two parameters: the midplane radius of the inner edge $r=r_{\mathrm{in}}$ and the radius of the maximum pressure at $r=r_{\mathrm{max}}$. See Appendix~\ref{app:fmtorus} for more detail.

\vspace{1em} 

\iharm is configured to run both SANE and MAD simulations and can be extended to support other initial conditions. Although the steady-state magnetic flux is not trivially related to the initial conditions of the simulation, we have implemented configurations that have been identified in previous work and shown to produce either a SANE or MAD flow. The initial condition for the magnetic field is determined by a prescribed, axisymmetric electromagnetic vector potential $A_\phi(r, \phi)$, which is computed at simulation zone corners. The magnetic field is then calculated from the curl of the vector potential using a finite different operator that is compatible with the constrained transport scheme to ensure that the magnetic field obeys the no-monopoles constraint (see \citealt{zilhao_2014_DynamicFisheyeGrids} for more detail).

The initial conditions are as follows: for SANE disks,
\begin{align}
    A_\phi = \mathrm{max}\left[\dfrac{\rho}{\rho_{\mathrm{max}}} - 0.2, 0\right],
\end{align} 
where $\rho_{\mathrm{max}}$ is the maximum initial plasma density; and for MAD disks,
\begin{align}
    A_{\phi} = \mathrm{max}\left[\dfrac{\rho}{\rho_{\mathrm{max}}} \left( \dfrac{r}{r_0} \sin \theta \right)^3 e^{-r/400} - 0.2, 0 \right],
\end{align}
where $r_0$ is chosen to be the inner boundary of the simulation domain  (see \citealt{hawley_2000_GlobalMagnetohydrodynamicalSimulations,gammie_2003_harm,wong_2021_JetDiskBoundaryLayer}). 
The MAD prescription concentrates the initial magnetic field toward the inner edge of the disk and forces it to taper at large $r$. In both the SANE and MAD cases, the magnetic field is normalized so that the ratio between the maximum $P_{\rm gas}$ and the maximum $P_{\rm mag} = b^2/2$ is some target value, typically $100$. See Appendix~\ref{app:fmtorus} for more details regarding our fluid initial condition.

Note that other initial conditions are possible, including alternative fluid tori solutions \citep{fishbone_1977_RelativisticFluidDisks, abramowicz_1978_RelativisticAccretingDisks,kozlowski_1978_AnalyticTheoryFluid,chakrabarti_1985_thickdisk,penna_2013_NewEquilibriumTorus} as well as implementations of more realistic ``boundary conditions'' determined by, e.g., simulations of magnetized stellar winds near the putative supermassive black hole in the Galactic Center \citep{ressler_2018_HydrodynamicSimulationsInner,ressler_2020_SurprisinglySmallImpact,ressler_2020_InitioHorizonscaleSimulations}.

\subsection{Ray-tracing images}

\ipole \citep{moscibrodzka_2018_ipole} is a publicly available GRRT code for covariant polarized radiative transfer and is a descendant of the unpolarized image code \ibothros \citep{noble_2007_ibothros}. \ipole produces a rectangular polarimetric image defined by a field of view (width in $GM/c^2$, or translated to $\mu$as in the context of EHT sources), distance from the black hole $d_{\mathrm{src}}$, and orientation with respect to the black hole spin axis and midplane (inclination and position angle). Each pixel reports
the specific intensities for the Stokes parameters $I_\nu, Q_\nu, U_\nu, V_\nu$ at the pixel center as well as the total optical and Faraday depths along the pixel-centered geodesic.

\ipole is an observer-to-emitter or backward-in-time code, meaning that only geodesics that intersect a predefined pinhole camera (observer) are considered. The camera is defined by a particular coordinate $x_c^\mu$ and an orthonormal tetrad specified by the normal observer velocity $u_\mu \propto \left(1, 0, 0, 0\right)$,\footnote{Earlier camera prescriptions defined the camera in the frame with $u^\mu \propto ( 1, 0, 0, 0)$, which is identical in the limit that the camera radius goes to infinity.} a radially directed wavevector, and the black hole spin axis. The pinhole camera defines an image with pixels that intersect $x_c^\mu$ and are regularly spaced in angle as measured in the tetrad frame. This prescription ensures that the central pixel corresponds to a geodesic with impact parameter zero. The geodesic for each pixel is integrated backward toward the black hole (i.e., the emitting matter) until it leaves the simulation's user-specified radiation domain, either at large radius or at the event horizon. The radiative transfer equation is then solved forward toward the camera.

\ipole uses operator splitting to solve Equation~\ref{eqn:covariantpoltransport} in two independent stages, which separately account for the relativistic parallel transport and nonrelativistic emission, absorption, and mixing radiative transfer updates. In the first substep, \ipole numerically integrates and parallel transports the coherency tensor $N^{\alpha\beta}$ along the geodesic. In the second substep, it projects $N^{\alpha\beta}$ into a Stokes vector $\boldsymbol{S}_\nu = \left(I_\nu, Q_\nu, U_\nu, V_\nu\right)$ in a local fluid tetrad,\footnote{Because we treat synchrotron radiation, it is convenient to construct the tetrad from the fluid $u^\mu$, the geodesic wavevector $k^\mu$, and the local magnetic-field orientation $b^\mu$.}  evaluates transfer coefficients in that tetrad, and uses a nonrelativistic analytic solution for the case of constant coefficients \citep{landideglinnocenti_1985_SolutionRadiativeTransfer} to perform transport in the tetrad frame. The use of an analytic solution ensures numerical stability when the plasma has large optical or Faraday depths. The code is coordinate agnostic, since all geometric factors are computed numerically from the metric. A comparison of contemporary GRRT codes is available in \citet{gold_2020_VerificationRadiativeTransfer}.

The Illinois version of \ipole, which is used in the \patoka pipeline, implements several new features and differs from the originally published version in several ways. First, Illinois \ipole implements an optional adaptive ray-tracing scheme that allows resolution to be concentrated in regions of the image with sharp, detailed features (for details see \citealt{gelles_2021_RoleAdaptiveRay}). The other differences are described below. Both the originally published version of \ipole and the Illinois version are publicly available.\footnote{The original version can be found on github at \linebreak \url{https://github.com/moscibrodzka/ipole}. The version described here can be found at \url{https://github.com/afd-illinois/ipole} as version 1.4. Both versions are currently maintained.} \ipole converges at second order.

\subsubsection{Tracking the emission source}

Since we aim to connect observables to the accretion flow structure and the spacetime geometry, it is useful to be able to study where the emission is produced. In order to compute what fraction of the emission produced by a given volume will contribute to an image, it is necessary to specify the location of the observer, since the emission coefficients depend on the photon wavevectors $k^\mu$ through the local sampling frequency $\nu= - k^\mu u_\mu$ and the pitch angle $=\arccos \left( k_\mu b^\mu \right)$. Furthermore, strong lensing allows multiple geodesics to sample the same region of space; although the photons may have the same frequency at infinity, in general each will be at a different frequency in the rest frame of the plasma since the geodesics are not parallel. Therefore a one-to-one mapping between points in space and emissivities does not exist.

Moreover, not all emitted light escapes to infinity. Light emitted near the event horizon may fall into the hole, and the optical depth of the plasma between emission and the observer will cause some light to be reabsorbed. It is straightforward to compute what fraction of light makes it to the observer at infinity for the Stokes $I$ total intensity by saving the values computed when solving the approximate radiative transfer equation, Equation~\ref{eqn:approxradxfer}, as follows.

At each step along the geodesic from $s_1 \to s_2$, we save the local optical depth and the local contribution to the invariant intensity, $\Delta \mathcal{I}_\nu= \mathcal{I}_\nu \left( s_2 \right) - \mathcal{I}_\nu \left( s_1 \right)$. After the full geodesic has been traced, the ``observed fraction'' of the photons emitted at any point along the ray that make it to the camera can be computed as $\exp\left( - \tau \right)$, where $\tau$ is the optical depth to the camera from that point. For each step along the geodesic, the local contribution to the final intensity is computed from $\Delta \mathcal{I}_\nu$ and the observed fraction; these values are then saved in an array representing the simulation domain. The process is validated after the complete image has been generated by summing the flux density contributed by each simulation zone and comparing to the final total image flux density. Since the output of this procedure is the total contribution from each simulation zone, it is important to account for the coordinate transformation Jacobian when plotting, e.g., the apparent density of the observed emitted photons.

\subsubsection{Subring decomposition}

Simulated images of EHT sources commonly display a clear ring-like structure. This structure is identified with the photon ring produced as strong lensing allows light paths to wind around the black hole multiple times (see, e.g., \citealt{johnson_2020_UniversalInterferometricSignatures}). In aggregate, the composition of the subrings produces a logarithmically divergent brightness temperature that is limited in part by the optical depth of the plasma.

Observable signatures of the photon ring structure have been studied analytically (e.g., \citealt{gralla_2020_LensingKerrBlack,himwich_2020_UniversalPolarimetricSignatures,chael_2021_ObservingInnerShadow,vincent_2021_GeometricModelingM87}), but the detailed structure of the observable photon ring is heavily influenced by the structure of the emitting plasma. We have included a subring decomposition routine in \ipole that allows the code to produce separate images for each subring. The subring structure is particularly evident when the emission exhibits a nontrivial latitudinal structure, e.g., if the emission is concentrated near the midplane and each subring contribution can be easily separated from its neighbors.

Geodesics that pass close to the hole experience strong lensing and may undergo latitudinal oscillations---this effect is most obvious for the unstable, constant-radius bound orbits, which circle the hole indefinitely and undergo periodic oscillations as they sweep through a range of latitudes \citep[see][]{teo_2003_SphericalPhotonOrbits}. Each event $x^\mu$ along a nonequatorial geodesic can be assigned an orbit number $n$, corresponding to the number of latitudinal turning points between the event and the camera in the region near where the geodesic passes through the midplane.\footnote{We have chosen for orbits to be defined with respect to $\theta$ rather than $\phi$. This choice leads to a subring index that roughly corresponds to the number of times the geodesic has passed through the midplane, although here $n$ may also increase as the geodesic tracks to infinity.} This value can be directly tracked and saved during the backward geodesic integration. Note that the subrings are nested and overlap: the $n^{\mathrm{th}}$ subring lies within the $\left(n-1\right)^{\mathrm{th}}$, with different segments of the same geodesic that undergoes $m$ latitudinal oscillations contributing to the first $m+1$ subrings. The $n=0$ component covers the full image, and higher-order subrings have exponentially decreasing areas as they tighten around the critical curve.

In order to synthesize the $n_{\mathrm{target}}^\mathrm{th}$ subring image, the emission coefficients are zeroed during the forward radiative transfer integration in regions of the geodesic with $n \neq n_{\mathrm{target}}$. It is important to include parallel transport as well as absorption and rotation in regions with $n \neq n_{\mathrm{target}}$, since although the subring image comprises only photons that were emitted along the corresponding geodesic segment, we must account for how those photons interact with matter and spacetime as they propagate to the observer. Thus, although $j_{I,Q,U,V}$ are set to zero, $\alpha_{I,Q,U,V}$ and $\rho_{Q,U,V}$ are not.

\subsubsection{Differences from the original version}

We have slightly modified \ipole from the originally published version in several ways:

\begin{enumerate}
    \item \ipole has been modified to read fluid snapshot files with arbitrary logical coordinate systems through the use of a new module called {\tt{}simcoords}. Using {\tt{}simcoords}, \ipole performs all ray-tracing in exponential Kerr--Schild (eKS; see Appendix~\ref{app:fmkscoords}) coordinates and uses an interpolated grid map between eKS coordinates and the input snapshot coordinates. In order to use the {\tt{}simcoords} module, a fluid snapshot must report the location of each grid zone in Kerr--Schild (KS) coordinates, provide the velocity and magnetic-field vectors with KS components, and correspond to a single slice of constant KS time.
    \item Several different methods for treating the $\sigma$ magnetization cutoff have been implemented. The default version treats $\sigma$ as an interpolated scalar and simply zeros the transfer coefficients in regions with $\sigma > \sigma_{\mathrm{cutoff}}$ by setting the number density of electrons $n_e = 0$. Alternative methods include gradually suppressing $n_e$ according to a sigmoidal function of $\sigma$ or zeroing $n_e$ on a per-zone basis according to zone-centered values of $\sigma$. The choice of cutoff procedure may introduce image artifacts---see Appendix~\ref{app:sigmacutoff}.
    \item When evaluating fluid parameters along geodesics, we interpolate the scalars $n_e$, $u/\rho$, magnetic-field strength as well as the six primitive variables that describe the fluid velocity and magnetic-field orientation. This procedure ensures that temperatures, which are derived from $u / \rho$, remain reasonable and that the interpolation scheme does not lower the magnetic-field strength in regions where the magnetic field oscillates wildly, like near the jet--disk boundary. Then, by constructing $u^\mu$ and $b^\mu$ from the primitives, we can ensure $u^\mu u_\mu = -1$ and $u^\mu b_\mu = 0$. We have observed that interpolating the four-vector components directly tends to produce $\ge \mathcal{O}(1)$ deviations from these criteria within $\approx 2\ GM/c^2$ of the event horizon. Deviations have been observed to be catastrophic in some cases when the {\tt{}simcoords} module is used.
    \item During evaluation of the polarized transfer coefficients, \ipole enforces positivity constraints on transfer coefficients to ensure that $j_{\nu,I} \ge 0$ and $\alpha_{\nu,I} \ge 0$ and that $j_{\nu,I} \ge \sqrt{j_{\nu,Q}^2 + j_{\nu,U}^2+j_{\nu,V}^2}$ and $\alpha_{\nu,I} \ge \sqrt{\alpha_{\nu,Q}^2 + \alpha_{\nu,U}^2+\alpha_{\nu,V}^2}$. Failure to enforce the latter condition leads to division-by-zero errors in evaluation of the analytic, constant-transfer-coefficients solution that \ipole uses to advance the Stokes vector. Failure to enforce the condition on $j_{\nu,I}$ can lead to polarization fractions $> 100\%$, and failure to enforce positivity on $j_{\nu,I}$ leads to unphysical negative intensities. Furthermore, \ipole reorthogonalizes the fluid-frame tetrads to ensure orthonormality to machine precision. Finally, consistent limiting expressions in areas of low absorption or rotation improve the accuracy of maps of polarization fraction.
    \item \ipole includes implementations of transfer coefficients for various nonthermal eDFs, such as the power law and kappa distributions. The coefficients are computed using \symphony~\citep{pandya_2016_PolarizedSynchrotronEmissivities,pandya_2018_NumericalEvaluationRelativistic,marszewski_2021_UpdatedTransferCoefficients}, which is now a submodule of the code.
    \item Several analytic and test models have been added to check \ipole against existing radiative transport schemes (see, e.g., \citealt{gold_2020_VerificationRadiativeTransfer} 
    and Appendix~\ref{app:polcomparison}).
    \item The coefficients of the radiative transfer equation, along with the fluid parameters that produced them, can be recorded along the geodesic corresponding to any pixel in the image and output as a \emph{trace} file. This enables easier \emph{post-hoc} study of regions that may produce characteristic emission and also allows for easier error diagnosis.
\end{enumerate}

\subsection{Producing spectra}

\igrmonty \citep{dolence_2009_grmonty} is a Monte Carlo radiative transport code designed to generate spectra from GRMHD fluid simulation snapshot files of optically thin ionized plasmas. It accounts for the full angle- and frequency-dependence of emission and absorption, and it treats single Compton scattering exactly. \igrmonty is an emitter-to-observer code, also known as forward-in-time, meaning it simulates emission at all frequencies and angles across the entire domain. This procedure is slower than the observer-to-emitter procedure, but it produces full spectra $\nu L_\nu$ of the source as seen from all inclinations and longitudes around the black hole. 

\igrmonty tracks a Monte Carlo sample of the radiation field in the form of \emph{superphotons}. A superphoton with weight $w$ is a packet of $w \gg 1$ photons, where each photon has the same position $x^\mu$ and wavevector $k^\mu$. Superphotons are created across the computational domain according to the local emissivity of the plasma. In order to emit a target number $N_{\mathrm{target}}$ of superphotons over the full simulation domain, the bolometric luminosity due to emission is precomputed for each zone and used to determine what fraction of the $N_{\mathrm{target}}$ total superphotons should be emitted per zone. This heuristic can produce a noisy signal when the fluid simulation resolution is increased if there is nontrivial structure in the accretion flow, thus higher fluid resolutions typically require a larger number of superphotons during the spectrum-generation step.

To optimize signal in the spectrum, the weight factor $w$ is conventionally chosen to be frequency-dependent according to the heuristic that each logarithmic bin in energy space in the final output spectrum should contain approximately the same number of superphotons. Since it is impossible to predetermine how many superphotons survive extinction on their journeys to the observer, we estimate how the precomputed weight factors should be set by assuming that all emitted photons escape to infinity---note that we relax this assumption when simultaneously treating both bremsstrahlung and synchrotron, as the two components tend to be peak at significantly different frequencies and thus undergo different amounts of extinction. We also neglect factors like redshift, scattering, and the ultimate angular dependence of the spectrum. 
When a new superphoton is created, its frequency is chosen according to rejection sampling and the rest of its wavevector is initialized in a local orthonormal tetrad according to the pitch-angle-dependent emissivity prescription. Each superphoton saves its initial position; this information can be used to infer properties of the emission region.

After emission, the optical depths to both absorption and scattering are computed as the superphotons propagate along their geodesics. Absorption is accounted for by decreasing the weight $w$ of the superphoton packet, which is directly proportional to the intensity of the radiation packet. 
\igrmonty treats scattering in a probabilistic sense: a superphoton will scatter with some probability at each step along its geodesic; if it scatters, the wavevector of the scattered superphoton is evaluated in a local orthonormal tetrad and determined probabilistically from the differential electron scattering cross section. 

In order to boost the signal in the Compton-upscattered component of the spectrum, \igrmonty uses a variant of the bias method introduced in \citet{kahn_1950_RandomSamplingMonte}. Although the likelihood $p$ that a full superphoton will scatter may be small in the optically thin plasmas we consider, the bias procedure enables a fraction $1/b$ of a fiducial superphoton to scatter with probability $bp$. Here, $b$ is the value of the bias parameter, which can in general vary across different simulation zones and time steps. If the superphoton scatters, its weight is decreased to $1 - 1/b$ and a new superphoton representing the scattered component is created with weight $1/b$.

\igrmonty is publicly available,\footnote{\url{https://github.com/afd-illinois/igrmonty}} converges at second order in geodesic integration, and converges like $\sqrt{N_s}$ in the number of Monte Carlo radiation field samples. The \patoka version of \igrmonty only solves the approximate (unpolarized) radiative transfer equation, but recently \citet{moscibrodzka_2020_polmonty} introduced the {\tt{}RADPOL} scheme, which accounts for fully polarized synchrotron emission, absorption, Faraday rotation and conversion, and Compton scattering. Appendix~\ref{app:polcomparison} presents a comparison between \igrmonty and \ipole in the lower-frequency part of the spectrum where Compton upscattering is unimportant. 

\subsubsection{Scattering bias factor}

It is difficult to determine how the bias should be set before running the simulation. If it is too low, too few superphotons will scatter and the Compton contribution to the spectrum will be unusably noisy. If it is too high, the code could reach a \emph{supercritical} state, in which superphotons produced through scattering also undergo scattering events and the total number of superphotons diverges. In all cases, the bias factor must be $\ge 1$.

The bias parameter is a function of the local fluid state and typically scaled by the square of the local electron temperature $\Theta_e^2$ to improve resolution in the scattering events with higher energies compared to the average. This location-dependent value is multiplied by a global tuning factor $b_{\mathrm{tuning}}$, which scales the scattering rate across the entire simulation. \igrmonty begins each run with a low-resolution ``bias tuning'' step, during which $b_{\mathrm{tuning}}$ is varied until the ratio between the number of superphotons created through scattering and the number created through emission is approximately unity. The ratio is tracked during the tuning runs, and the evaluation is halted if the ratio exceeds a large number to ensure that the supercritical state can be preempted.

\subsubsection{Bremsstrahlung}

\citet[][see also \citealt{narayan_1995_ADAF,esin_1996_HotOneTemperatureAccretion,mahadevan_1997_ScalingLawsAdvectiondominated}]{yarza_2020_BremsstrahlungGRMHDModels} found that, in radiative simulations of SANE accretion flows, bremsstrahlung emission may dominate the high-energy ($\gtrsim~512$~keV) component of the spectrum and contribute to the bolometric luminosity as the accretion rate $\dot{M}$ is increased. In relativistic plasmas, both electron--electron and electron--ion bremsstrahlung contribute to the total emissivity; \igrmonty supports several contemporary prescriptions for both emissivities (see the Appendix of \citealt{yarza_2020_BremsstrahlungGRMHDModels} for a comparison of the prescriptions).
Since synchrotron and bremsstrahlung emission may be simultaneously nonzero, each emitted superphoton is assigned a bremsstrahlung-fraction value between zero and one. This value is saved when the spectrum is recorded and can be used to determine how much emission is associated with which emission process. 

Bremsstrahlung emission dominates direct synchrotron at higher frequencies where the optical depth of the plasma is low. A naive application of the superphoton weight assignment scheme described above may thus preferentially improve the signal in the bremsstrahlung component of the spectrum. This shortcoming can be addressed by either of the following: modifying the weighting procedure to independently generate weights for the lower-frequency synchrotron emission and the higher-frequency bremsstrahlung emission; or independently generating spectra from the synchrotron- and bremsstrahlung-emission components and adding them together. Compton scattering is typically unimportant for bremsstrahlung emission, so the bremsstrahlung calculation may be performed with scattering turned off. 

\subsubsection{Arbitrary electron distribution functions}

\igrmonty relies on the full eDF both when computing the Compton scattering cross section and during the scattering procedure when choosing an electron off which to scatter. The version of the code published in \citet{dolence_2009_grmonty} implemented a semianalytic treatment of the cross section and scattering sampling routines; here, we instead use an analytic expression only for the distribution function itself and numerically evaluate the cross section. We use the rejection sampling technique to sample the distribution rather than rely on an inversion of the distribution function. Because this procedure relies only on a prescription for the distribution function, it is quite general and supports any isotropic distribution that can be written as a function of the local fluid parameters.

\section{Sample Simulation Data Products}
\label{sec:results}

\begin{deluxetable*}{ lrllllcc }
\tablecaption{List of GRMHD Simulations} \label{table:grmhd_models}
\tablehead{ 
\colhead{flux} & 
\colhead{$\bhspin$} &
\colhead{$r_{\mathrm{in}}$} &
\colhead{$r_{\mathrm{max}}$} &
\colhead{$r_{\mathrm{out}}$} &
\colhead{$\hat{\gamma}$} &
\colhead{resolution(s)} &
\colhead{notes} 
}
\startdata
SANE & $0.94$ & $6$ & $12$ & $50$ & 4/3 & 192x192x192, 288 & 288 canonical \\
SANE & $0.75$ & $6$ & $12$ & $50$ & 4/3 & 192, 288 & -- \\
SANE & $0.5$ & $6$ & $12$ & $50$ & 4/3 & 192x192x192, 288 & 288 canonical \\
SANE & $0$ & $6$ & $14$ & $50$ & 4/3 & 288 & canonical \\
SANE & $-0.25$ & $8$ & $17$ & $50$ & 13/9 & 288 & -- \\
SANE & $-0.5$ & $8$ & $17$ & $50$ & 4/3 & 288 & canonical \\
SANE & $-0.5$ & $10$ & $20$ & $50$ & 5/3 & 192x128x128, 288x192x192 & -- \\
SANE & $-0.75$ & $10$ & $20$ & $50$ & 13/9 & 288 & -- \\
SANE & $-0.94$ & $10$ & $20$ & $50$ & 4/3 & 288 & canonical \\ 
SANE & $-0.94$ & $10$ & $20$ & $50$ & 5/3 & 192x128x128, 288x192x192 & -- \\ \hline
MAD & $0.94$ & 20 & 41 & $1000$ & 5/3 & 288, 384 & -- \\
MAD & $0.94$ & 20 & 41 & $1000$ & 13/9 & 192, 288, 384, 448 & 384 canonical \\
MAD & $0.75$ & 20 & 41 & $1000$ & 13/9 & 384 & -- \\
MAD & $0.5$ & 20 & 41 & $1000$ & 13/9 & 192, 288, 384, 448 & 384 canonical \\
MAD & $0.25$ & 20 & 41 & $1000$ & 13/9 & 192, 384 & -- \\
MAD & $0$ & 20 & 41 & $1000$ & 13/9 & 192, 288, 384, 448 & 384 canonical \\
MAD & $-0.25$ & 20 & 41 & $1000$ & 13/9 & 192, 384 & -- \\
MAD & $-0.5$ & 20 & 41 & $1000$ & 13/9 & 192, 288, 384, 448 & 384 canonical \\
MAD & $-0.94$ & 20 & 41 & $1000$ & 13/9 & 192, 288, 384, 448 & 384 canonical \\
\enddata
\tablecomments{
GRMHD fluid simulation parameters. Canonical simulations are identified in the notes column. Flux labels the relative strength of the magnetic flux at the horizon, $\bhspin$ describes the spin of the black hole, $r_{\mathrm{in}}$ and $r_{\mathrm{max}}$ are parameters for the initial Fishbone--Moncrief torus, $r_{\mathrm{out}}$ is the outer edge of the simulation domain, $\hat{\gamma}$ is the adiabatic index of the fluid, and resolution gives the $N_r \times N_\theta \times N_\phi$ number of grid zones in the simulation. 
In the case of resolution, single-number abbreviations mean the following: 192 $\to$ 192x96x96, 288 $\to$ 288x128x128, 384 $\to$ 384x192x192, and 448 $\to$ 448x224x224.
}
\end{deluxetable*}

\begin{figure*}
\centering
\includegraphics[trim = 0.5cm 0cm 0.5cm 0mm,width=0.95\textwidth]{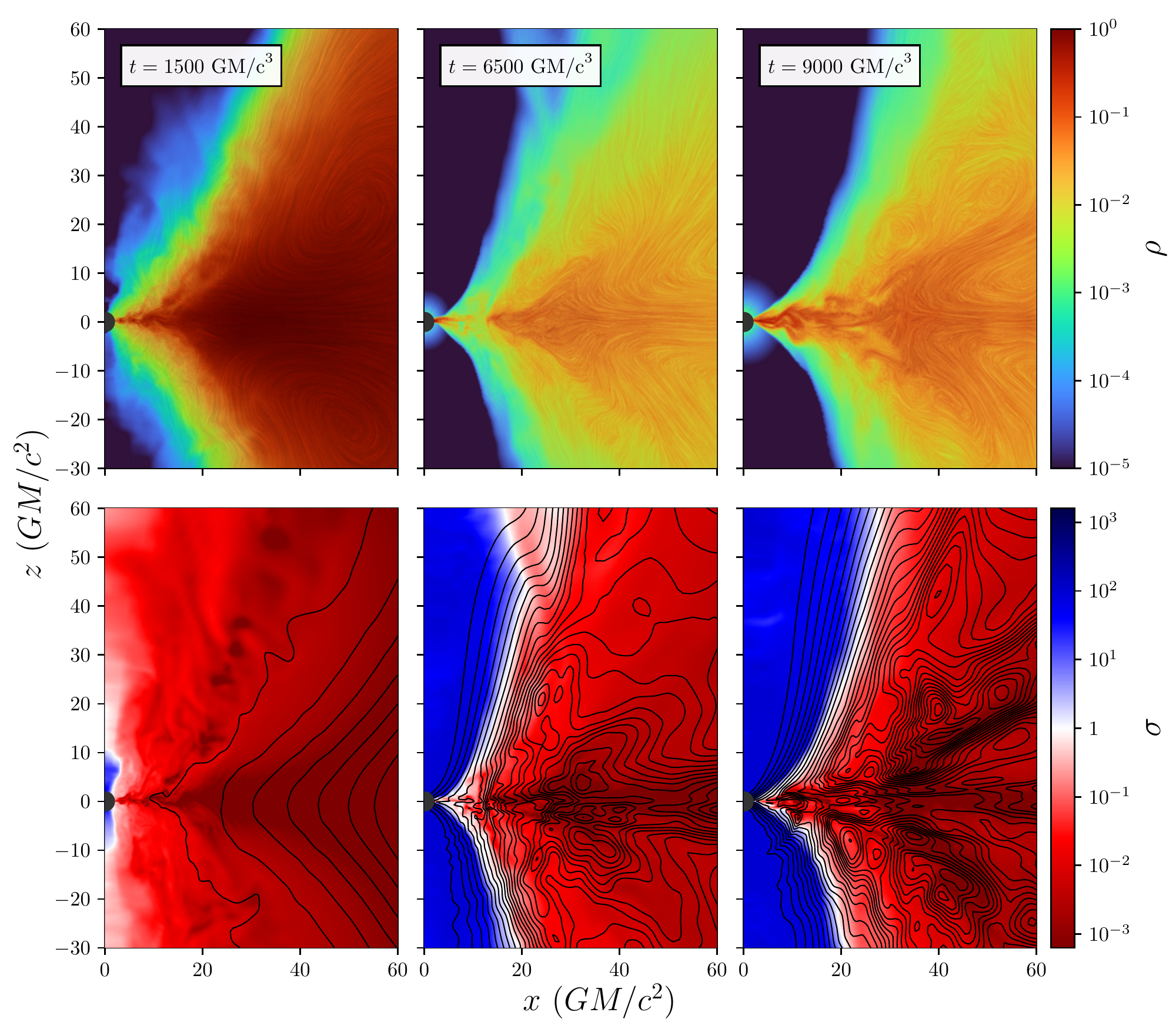}
\caption{GRMHD simulation of a MAD accretion flow with $\bhspin = 0.5$. This simulation was initialized with an $r_{\mathrm{in}}=20$, $r_{\mathrm{max}}=41$ torus and run at a resolution of $448 \times 224 \times 224$ zones in the $r$, $\theta$, and $\phi$ directions, respectively. Top panels show rest-frame plasma density in arbitrary units at three different times over the course of the simulation; the line integral convolution technique is used to represent the motion of the plasma. The bottom panels show magnetization at the same times as the top panels with overplotted contours of the axisymmetrized vector potential component $A_\phi$, which give a sense of field strength and disorder. The black hole event horizon is plotted as a dark circle. As the simulation evolves, the flow becomes increasingly turbulent and a high-magnetization jet region opens around the poles.}
\label{fig:grmhd_visual_evolution_mad}
\end{figure*}

\begin{figure*}
\centering
\includegraphics[trim = 0.5cm 0cm 0.5cm 0mm,width=0.75\linewidth]{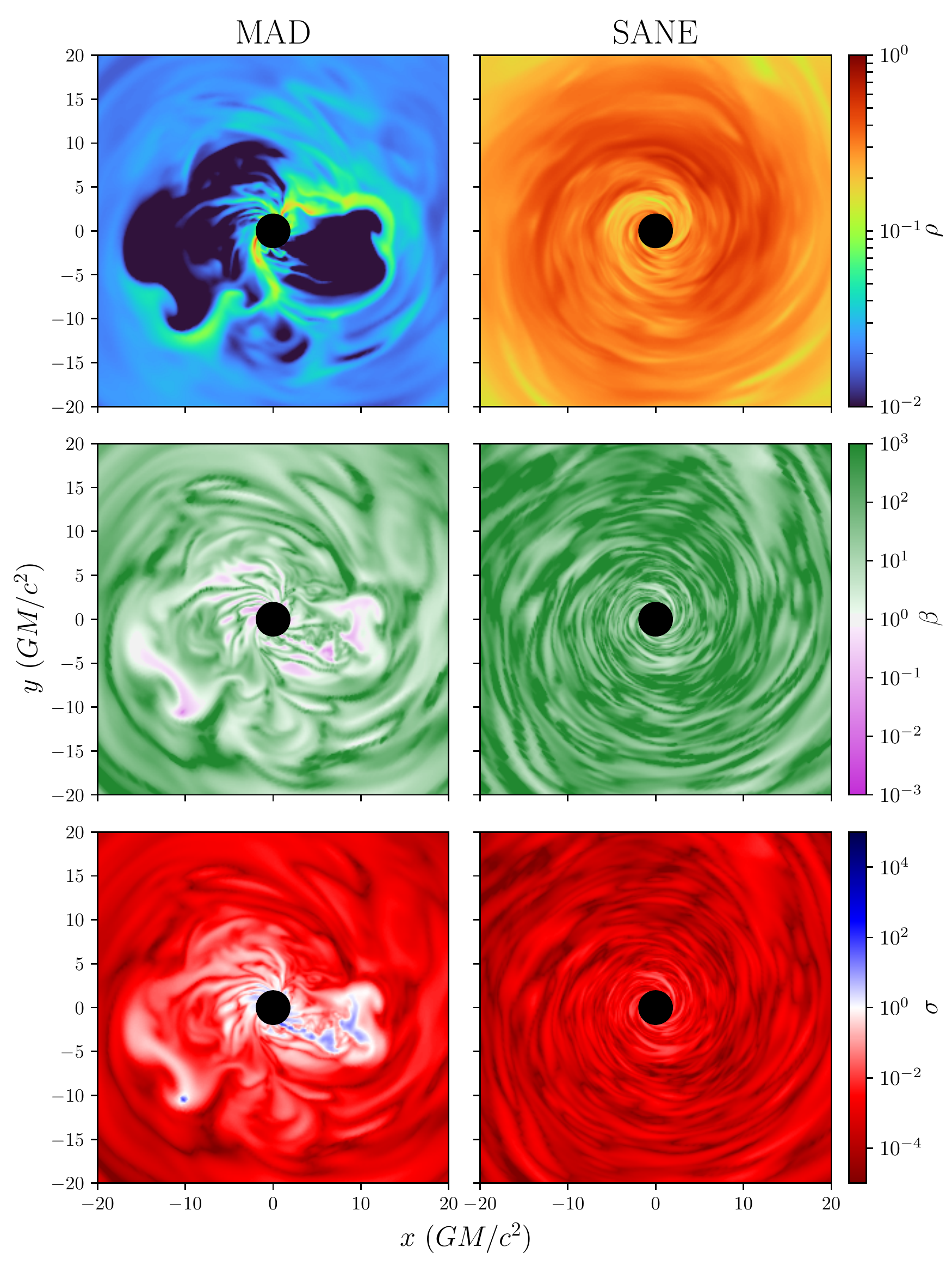}
\caption{Comparison of density, plasma $\beta$, and magnetization $\sigma$ in the midplane of MAD and SANE simulations in steady state. Both simulations were run with an intermediate spin $\bhspin = 0.5$. In the MAD simulation, the accretion proceeds in thin strands, contrasting the steady but turbulent disk-like SANE accretion mode. The evacuated regions in the MAD simulation with low $\beta$ and high $\sigma$ are magnetic bubbles produced during the flux ejection events that episodically recur when an excess of magnetic flux has been trapped on the event horizon.}
\label{fig:topdown_madsane}
\end{figure*}

We now present example data products from the \patoka pipeline. The data we show here were generated for the EHT M87 simulation library that was used for validation in \citetalias{EHTC_2019_4,EHTC_2021_7} and for analysis in \citetalias{EHTC_2019_5,EHTC_2019_6,EHTC_2021_8}. We also present examples of unpublished data that were produced for \citetalias{EHTC_2019_4,EHTC_2019_5,EHTC_2019_6,EHTC_2021_7,EHTC_2021_8}.

\subsection{GRMHD models}
\label{sec:grmhd_models_and_floors}

The GRMHD library generated by \patoka for the EHT M87 analysis included two parts. The first part comprised ten ``canonical'' simulations spanning the MAD and SANE accretion states and five spins $\bhspin = -15/16, -1/2, 0, +1/2, +15/16$.\footnote{Here, negative values of spin imply that the angular momentum of the hole and the disk are antiparallel, i.e., the system is \emph{retrograde}.} Hereafter, we write $1/2$ as $0.5$ and $15/16$ as $0.94$ to be consistent with the EHT paper sequence. 
The second part was composed of several dozen ancillary simulations used to test different initial conditions, disk sizes, adiabatic indices, and simulation resolutions.
Table~\ref{table:grmhd_models} lists all simulations and identifies which simulations belong to the canonical set.
Each of the models generated for the library was evolved until at least $t = 10^4\ GM/c^3$, during which time its accretion flow reaches a statistical steady state within $r \le 10-20\ GM/c^2$. GRMHD fluid snapshots were saved every $5\ GM/c^3$, corresponding to $\approx 43$ hours for a black hole with mass $\approx 6.2 \times 10^9\ M_\odot$.

The following set of limiting conditions was imposed on the fluid evolution to ensure stability:
\begin{itemize}
    \item density $\rho > 10^{-6}\, k$ for $k \equiv \frac{1}{r^2 (1 + r/10)}$,
    \item internal energy $u > 10^{-8} \, k^{\hat{\gamma}}$,
    \item $\rho$ and $u$ were increased until $\sigma < 400$ and $\beta \equiv \frac{P_{\mathrm{gas}}}{P_{\mathrm{mag}}} > 2.5 \times 10^{-5}$, where $P_{\mathrm{mag}} = \frac{b^2}{2}$ is the magnetic pressure,
    \item $\rho$ was increased until $\frac{u}{\rho} < 100$,
    \item when evolving electron temperatures, $u$ was decremented until $\frac{P_{\mathrm{gas}}}{\rho^{\hat{\gamma}}} < 3$, and
    \item the velocity components were downscaled until the Lorentz factor with respect to the normal observer was $\Gamma = - u^\mu n_\mu < 50$.
\end{itemize}
Global disk simulations inevitably invoke some of these bounds, most frequently the ones on $\sigma$ and $\Gamma$. The former is visible in the top-right panel of Figure \ref{fig:grmhd_visual_evolution_mad} as a halo of accretion near the event horizon.

We only produced simulations with purely corotating (aligned) or counterrotating (antialigned) accretion flows, since varying disk tilt adds another dimension to the parameter space and is thus prohibitively expensive. We only produced SANE simulations with $\phi \approx 1$, although initial conditions that produce $\phi \ll 1$ and $1 \lesssim \phi \lesssim \phi_c$ accretion states are also known.
The fluid was assumed to have a uniform constant adiabatic index for each simulation $\hat{\gamma}$, although the value of $\hat{\gamma}$ was varied between different simulations.

Each simulation was run on a three-dimensional regular grid defined in the horizon-penetrating funky modified Kerr--Schild (FMKS; see Appendix~\ref{app:fmkscoords}) coordinates, which concentrate resolution toward the midplane and away from the jet at small radius. The simulation domain extended from $\ge 5$ simulation zones within the event horizon to an outer radius $r_{\mathrm{out}} \ge 50 GM/c^2$, depending on the simulation. 
We found that evolving a large disk over a long time could cause an initially SANE accretion flow to go MAD, so we chose to initialize the canonical SANEs with smaller accretion disks, allowing us to use smaller simulation domains and lower absolute resolutions.

Figure~\ref{fig:grmhd_visual_evolution_mad} shows snapshots of the plasma rest-mass density $\rho$ and magnetization $\sigma$ over the course of evolution of an intermediate-spin $\bhspin = 0.5$ MAD simulation. The initial condition with a large Fishbone--Moncrief torus and an ordered, looping magnetic field, progressively gives way to a turbulent accretion flow with a low-density high-magnetization funnel containing a strong, ordered magnetic field. The qualitative difference between the MAD and SANE accretion morphologies is shown in Figure~\ref{fig:topdown_madsane}. The MAD simulation is punctuated by magnetic bubbles corresponding to flux ejection events.

\begin{figure*}[htp!]
\centering
\includegraphics[width=\linewidth]{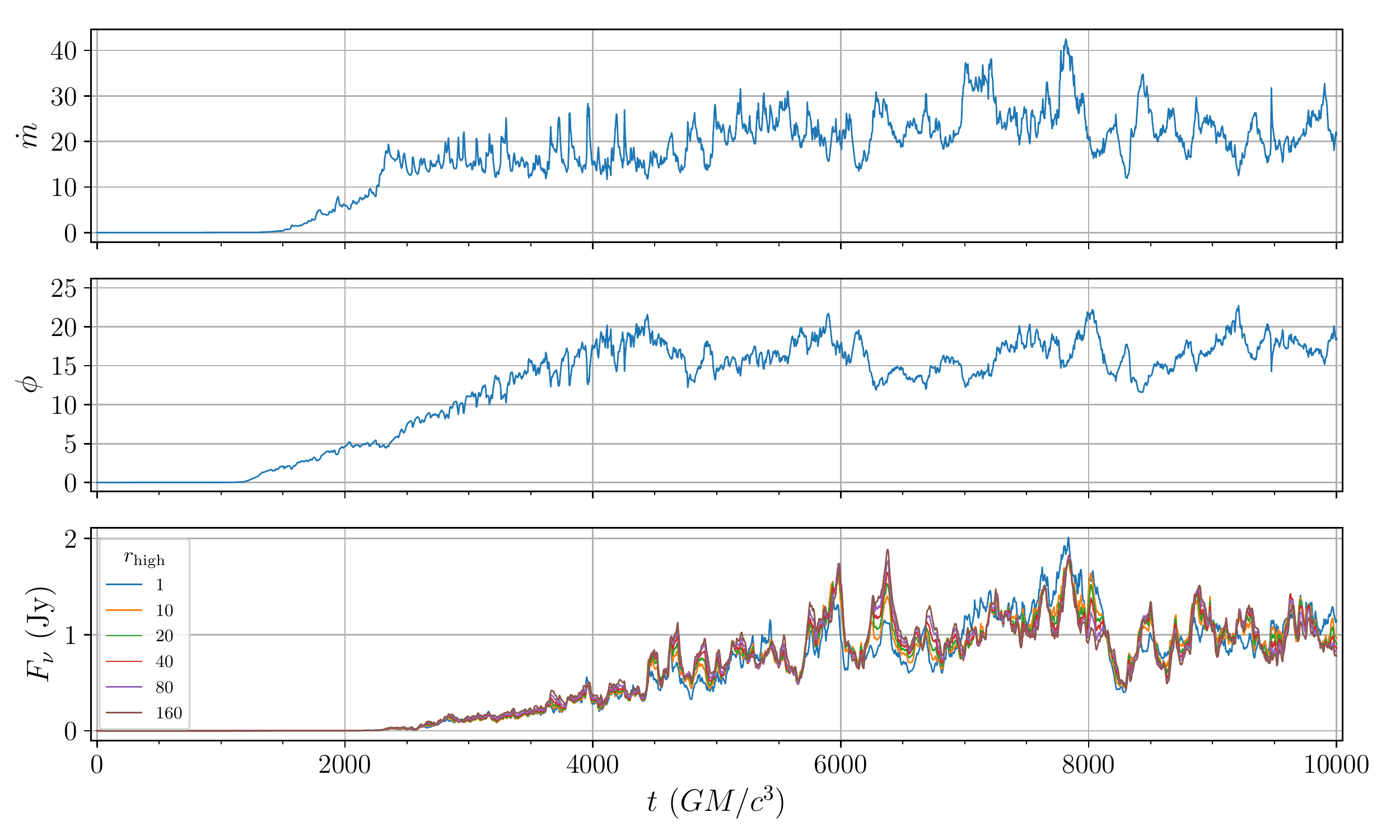}
\caption{Time series of flux variables from the MAD $\bhspin = 0.5$ simulation. Top panel: mass accretion rate in arbitrary code units versus time. Center panel: dimensionless magnetic flux $\phi$ versus time. Bottom panel: Light curves at $\nu = 230\,$GHz for $r_{\mathrm{low}} = 1$ and different $r_{\mathrm{high}}$ values. The light curves have been scaled to match a $1$ Jy target flux density over the last $5000\ GM/c^3$ of the simulation. The first half of the simulation is dominated by a transient from the torus initial condition. Notice that stability in $\dot{m}$ and $\phi$ does not necessarily equate to stability in the $1.3$ mm light curve. }
\label{fig:grmhd_fluxes}
\end{figure*}

\begin{figure*}[htp!]
\centering
\includegraphics[width=0.98\linewidth]{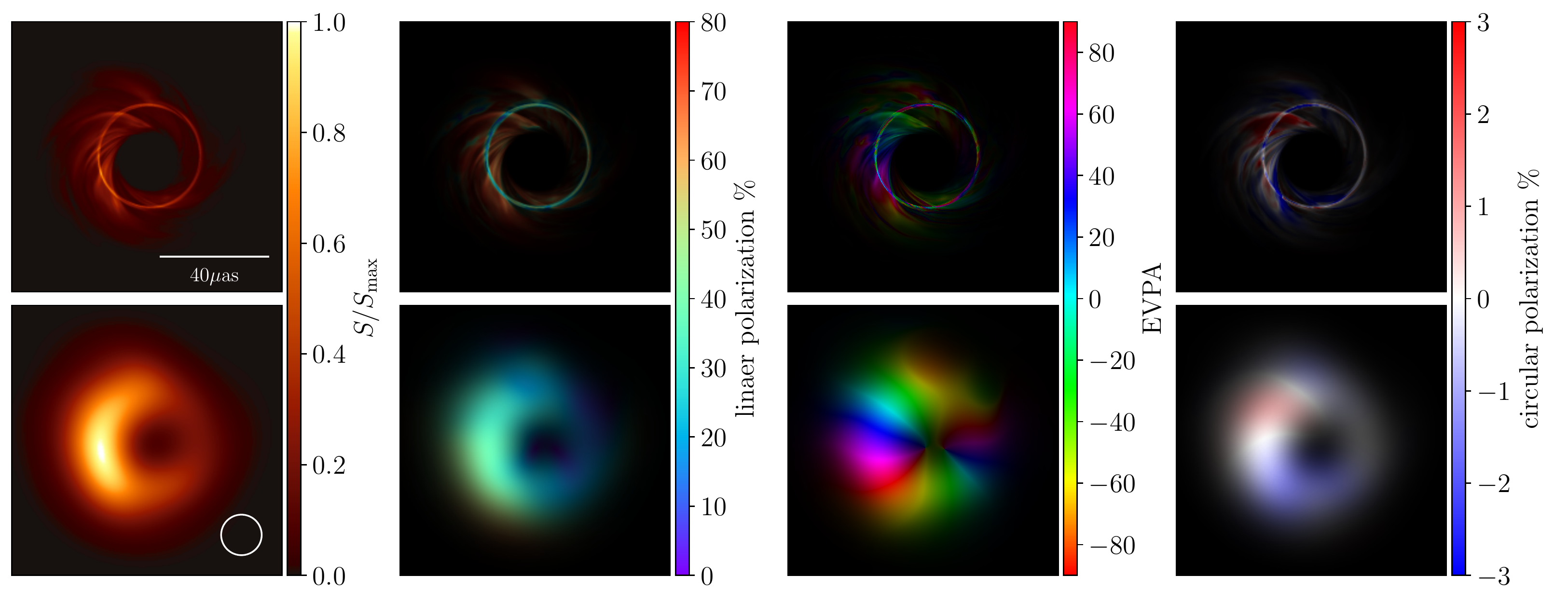}
\caption{
Example polarimetric images of MAD $\bhspin=0.5$ simulation produced by \ipole with (bottom) and without (top) blurring to $15\ \mu$as Gaussian beam. Panels show total intensity, linear polarization fraction, electric-vector position angle (EVPA), and circular polarization fraction. The brightness of each pixel has been scaled to the total intensity; the patterns in linear polarization, EVPA, and circular polarization continue into the dark regions. Blurring decreases the observed linear polarization fraction in regions where EVPA is rapidly varying.
}
\label{fig:grrt_example}
\end{figure*}

\subsection{Electromagnetic observables}
\label{sec:em_obs}

The GRMHD simulations were postprocessed to generate images and spectra using \ipole and \igrmonty. In total, about $120000$ images were generated for the canonical M87 total intensity and polarization analyses, and about $3$ million were generated for supporting analyses, including resolution and field-of-view studies, explorations of the analyses robustness to changes in numerical parameters like geodesic step size, and machine-learning projects. The full set of radiative transport model parameters is described below. The canonical images for the total intensity analysis were generated at $160 \times 160$ pixels resolution, and the images generated for the polarization analysis were at $320 \times 320$ pixels. All images were produced to have a $160\,\mu$as field of view.

For the images produced for the EHT M87 sequence, we assume that the accreting plasma is composed of pure ionized hydrogen, so that $y=z=1$ in Equation~\ref{eqn:equationte}. We fixed $\beta_{\mathrm{crit}} = 1$ and allowed the two parameters in Equation~\ref{eqn:tptemodel} to vary between $r_{\mathrm{low}} =  1, 10$ and $r_{\mathrm{high}} = 1, 10, 20, 40, 80, 160$. See \citetalias{EHTC_2019_5} and \citetalias{EHTC_2021_8} for a discussion of the motivation behind these choices.

The library discussed here was generated to compare against observations of M87, so it was generated using physical parameters that would target that system.
The inclination angle $i$ was chosen to be consistent with the orientation of the M87 jet at large scales, $i \approx 17^\circ$ \citep{hada_2017_m87jet,kim_2018_LimbbrightenedJetM87,walker_2018_m87jet}, so we produced a library with inclinations ranging over multiple inclinations $i = 7, 12, 17, 22, 27$ degrees relative to the line of sight. We do not know \emph{a priori} whether the black hole spin axis is directed toward or away from us. An exploratory survey of the library showed that it was necessary to orient the black hole spin vector away from Earth in order to reproduce both the image brightness asymmetry and the position angle of the large scale jet \citepalias{EHTC_2019_5}. The position angle of the spin axis can be reoriented during analysis by rotating both the image and the per-pixel EVPA.

The GRMHD simulation mass density was scaled to physical units by requiring that the simulated compact flux density at $230\,$GHz be consistent with the observed contemporaneous flux density of between $0.5 - 0.7$ Jy---see \citetalias{EHTC_2019_4} for more detail on identifying this target flux density. The relationship between the scaling factor $\mathscr{M}$ and total flux density $F_{\mathrm{tot}}$ is a complicated function of the accretion flow, but it tends to be monotonic near the target value, so identifying the appropriate scaling factor corresponds to a simple root-finding procedure. Since the fitting procedure is expensive, the flux density is typically fit using the approximate total intensity solution over a regular subsample of the snapshots at low resolution. The quality of the fit is substantiated when the high-resolution data are generated.
After identifying the value of $\mathscr{M}$ required to produce the target flux density, every snapshot from each GRMHD model is typically imaged, producing a sequence with a $5\ GM/c^3$ cadence. 

The result of running the flux-density-fitting procedure is shown in Figure~\ref{fig:grmhd_fluxes} for the canonical MAD $\bhspin = 0.5$ GRMHD simulation with $r_{\mathrm{low}}=1$ and $i = 17^\circ$. The unscaled mass accretion rate of the system $\dot{m}$ and the dimensionless magnetic flux parameter $\phi$ are also plotted for comparison. Here, the light curves have been fit so that the average $230\,$GHz flux density matches $1$ Jy over the last $5000\ GM/c^3$ of the simulation. 

Figure~\ref{fig:grrt_example} shows an example M87-like synthetic image from the MAD $\bhspin = 0.5$ model with $r_{\mathrm{low}} = 1$ and $r_{\mathrm{high}} = 40$ at $i=17^\circ$. Each pixel contains the full polarimetric Stokes $I, Q, U, V$ specific intensities, which can be processed to provide information about the linear polarization fraction $\mathrm{LP} = \sqrt{Q^2 + U^2}/I$, EVPA $\chi$, and circular polarization fraction $\mathrm{CP} = V/I$. Blurring is performed by convolving the Stokes intensities with a $15\,\mu$as Gaussian beam. In the right three columns, pixel brightness is determined by total intensity so that low-intensity regions, which contribute little flux in an observation, appear black.

\begin{figure*}[htp!]
\centering
\includegraphics[trim = 0.5cm 0cm 0.5cm 0mm,width=\linewidth]{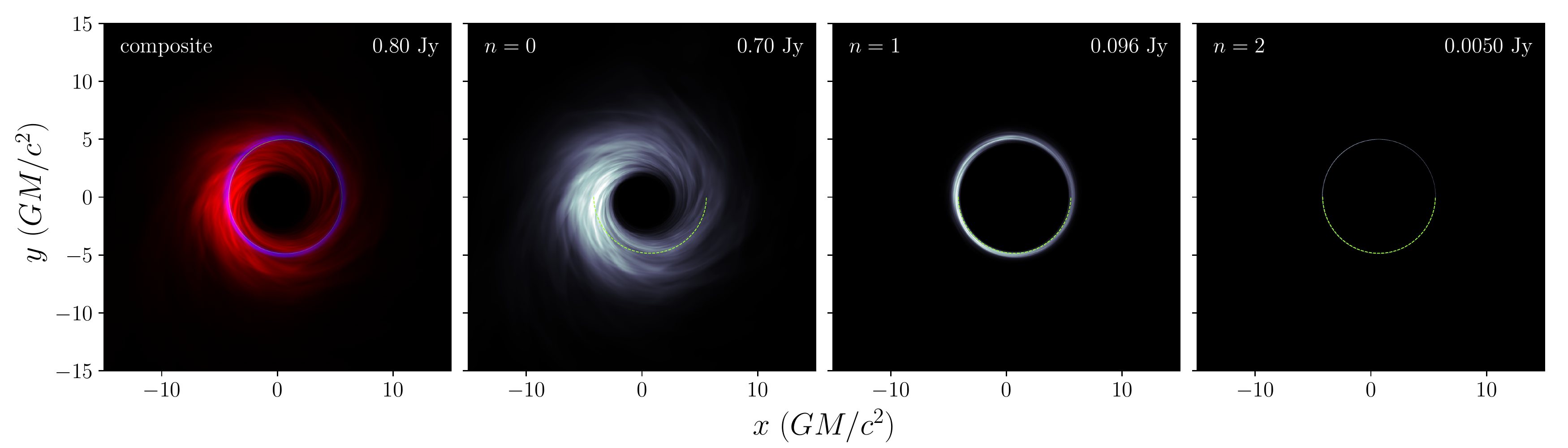}
\caption{Decomposition of the total intensity into the first three subrings for a snapshot from the high spin $\bhspin = 0.94$ MAD model with $r_{\mathrm{low}} = 1$, $r_{\mathrm{high}} = 10$, and $i = 17^\circ$. The intensities in the $n=0$, $n=1$, and $n=2$ subrings are reported in the right three panels, and the location of the critical curve is plotted as a dashed line in green. The total flux density, whose decrease is governed by angle-dependent Lyapunov exponents in the limit of large $n$ for optically thin systems, is given in the top right of each panel. The left panel shows a composite image produced by adding the $n=0$ subring in red, $n=1$ subring in blue, and $n=2$ subring in green. The visual appearance of each subring image is due to the details of the underlying emission, although the geometric extent of each image is due to the spacetime.}
\label{fig:grrt_ringdecomp}
\end{figure*}

Figure~\ref{fig:grrt_ringdecomp} shows the result of the ring decomposition procedure used to isolate the different subrings in a ray-traced image (see, e.g., Figure~3 of \citealt{johnson_2020_UniversalInterferometricSignatures} for cross sections of a similar decomposition). As $n$ increases, the extent of each next subring, which corresponds to a geometric region on the image, is exponentially demagnified compared to the last one according to Lyapunov exponents at each angle on the image (see \citealt{johnson_2020_UniversalInterferometricSignatures,wong_2021_BlackHoleGlimmer}). In physical systems, the increasing optical depth of the source limits how many subrings contribute to the composite image.

\begin{figure*}[htp!]
\centering
\includegraphics[trim = 0.5cm 0cm 0.5cm 0mm,width=\linewidth]{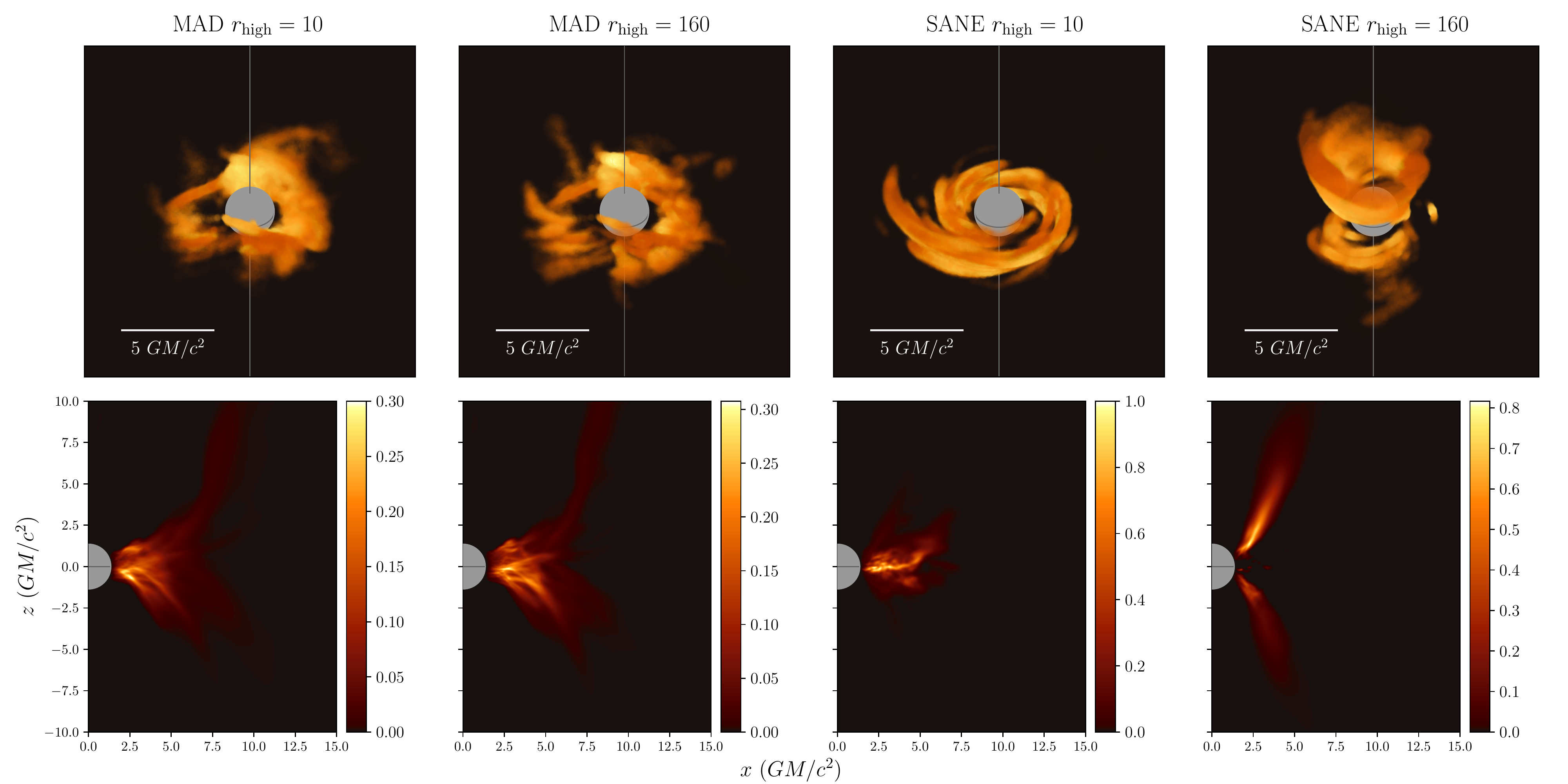}
\caption{Location of emission around high-spin black holes with $\bhspin = 0.94$ on a single KS time slice. In this figure, the emissivity is embedded in and represented with respect to a Cartesian coordinate system from simulation coordinates, and all densities are with respect to the Cartesian volume element. Top: three-dimensional rendering of emission source with color and transparency determined by the total emission produced within that region of space. Bottom: the same data as above after summing across the azimuthal dimension $\phi$. The left two columns are a representation of emission from a single snapshot from a typical MAD simulation; emission tracks the fragmentary plasma and is relatively insensitive to the electron thermodynamics, parameterized here by $r_{\mathrm{high}}$. The right two columns represent emission from the same SANE simulation snapshot but with the different electron temperature prescriptions. Larger values of $r_{\mathrm{high}}$ shift the emission from the turbulent but steady disk into the funnel region. All simulations have $r_{\mathrm{low}} = 1$ and are imaged at $i=17^\circ$. The total flux density produced by each simulation is the same, so the color scales show the relative concentration of emission in the azimuthal sum.
}
\label{fig:grrt_decomp}
\end{figure*}

\ipole can be used to track the source of the observed flux density, as seen in Figure~\ref{fig:grrt_decomp}. All MAD simulations tend to show the same characteristic fragmentary emission structure, which corresponds to the disjoint accretion (see, e.g., the top left panel of Figure~\ref{fig:topdown_madsane}). Much of the emission in the MAD case is thus produced in the hot, chaotic region of the flow near the horizon. SANE emission is comparatively more structured. Changing the electron temperature model in the SANE simulations can have drastic effects by heating up the jet funnel and shifting emission out of the disk.

\begin{figure*}[ht!]
\centering
\includegraphics[trim = 0.5cm 0cm 0.5cm 0mm,width=0.98\linewidth]{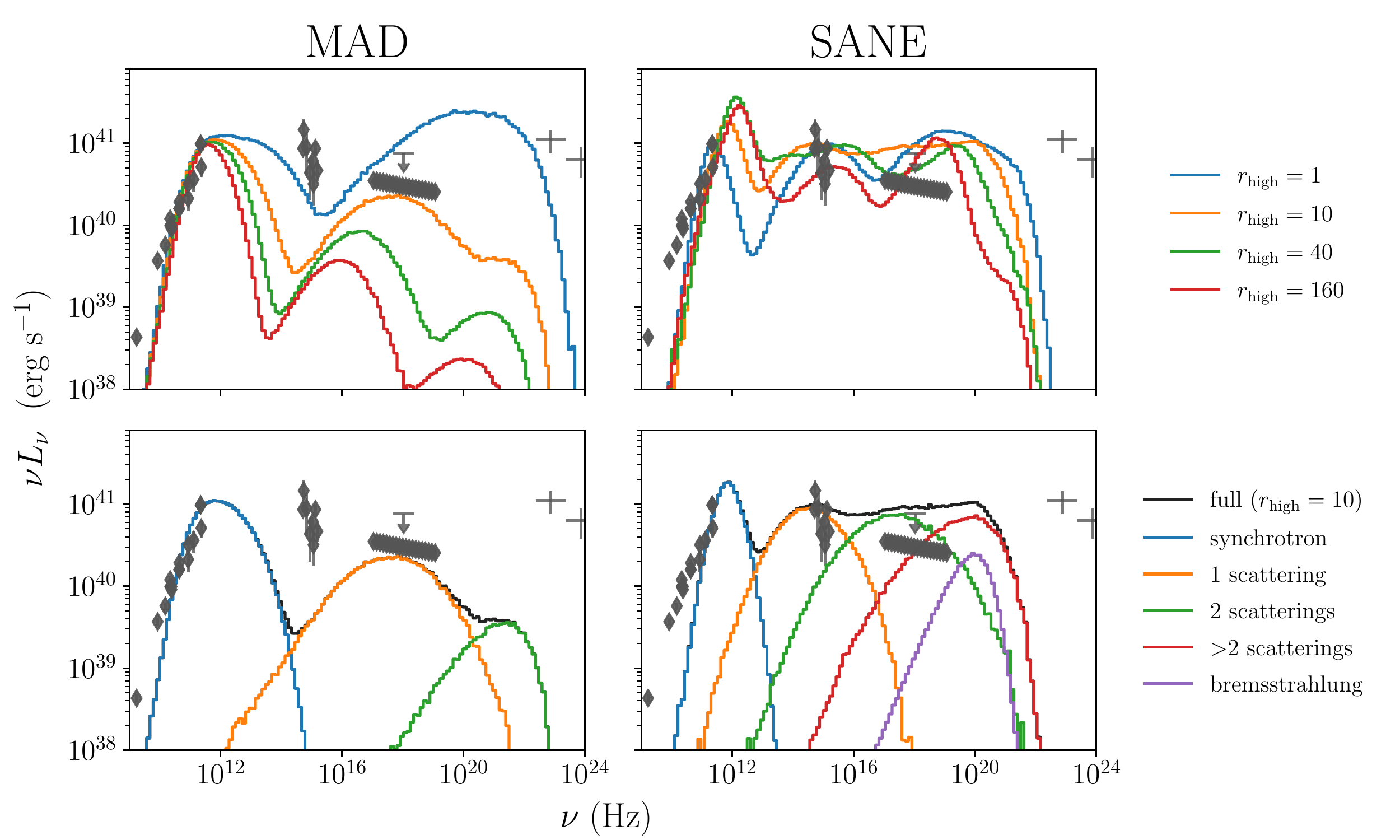}
\caption{Example spectra for MAD (left) and SANE (right) snapshots at low inclination in an $r_{\mathrm{low}}=1$ model. The spectra for $r_{\mathrm{high}}=1, 10, 40, 160$ are shown in the top row, and the components of the $r_{\mathrm{high}}=10$ model are shown in the bottom row. All data points are taken from the simultaneous multiwavelength measurement campaign performed coincident with the 2017 EHT observations of M87 and reported in \citet{ehtmwl_2021_BroadbandMultiwavelengthProperties}.}
\label{fig:grsm_example}
\end{figure*}

Running \igrmonty is significantly more computationally expensive than running \ipole, so it is infeasible to generate spectra for every fluid snapshot across every radiation model. The two spectrum constraints considered in the EHT analysis were the overall radiative efficiency of the flow and a boolean determination if the simulation X-ray flux was consistently too high. At minimum, we generate ten spectra per radiation model, but we checked that producing a denser sampling of spectra in time does not change the result in a statistical sense. Figure~\ref{fig:grsm_example} shows example spectra produced from one MAD and one SANE snapshot at $r_{\mathrm{low}} = 1$ and across different values of $r_{\mathrm{high}}$.

\section{Future directions}
\label{sec:discussion}

We now briefly discuss future directions as well as improvements and modifications that can be made to the \patoka pipeline.

\subsection{Radiative transfer model}

\ipole produces images calculated at a single frequency, which neglects the observing bandwidth of the instrument. Extensions to the ray-tracing code could allow for synthesis of finite bandwidth observations; however, this approximation has been found to be inconsequential in the context of the M87 library.

Our treatment generally assumes that the eDF is thermal, i.e., that it is well described by a Maxwell--J{\"u}ttner distribution. This assumption enters through definition of the transfer coefficients, which are calculated from the underlying distribution function (see, e.g., \citealt{shcherbakov_2008_PropagationEffectsMagnetized,pandya_2016_PolarizedSynchrotronEmissivities,pandya_2018_NumericalEvaluationRelativistic}). The introduction of nonthermal electrons can change both the spectral shape and the image morphology (see, e.g., \citealt{ozel_2000_HybridThermalNonthermalSynchrotron,yuan_2003_RIAFSgra,broderick_2009_ImagingBlackHole,chael_2017_EvolvingNonthermalElectrons,mao_2017_ImpactNonthermalElectrons,davelaar_2018_GeneralRelativisticMagnetohydrodynamical}).

GRMHD simulations produce snapshots of the fluid at different, discrete KS times. The \ipole and \igrmonty data we present were generated under the \emph{fast light} approximation, where only a single snapshot is used to generate an image or spectrum. This approximation is invalid if the fluid evolves on timescales shorter than the simulation light-crossing time or if one wishes to simulate various time-dependent phenomena, like light echoes from flares or glimmer (e.g., \citealt{broderick_2005_ImagingBrightspotsAccretion,moriyama_2015_NewMethodBlackhole,wong_2021_BlackHoleGlimmer,hadar_2021_PhotonRingAutocorrelations}).

The alternative \emph{slow light} method relies on a high fluid snapshot cadence, has large data-storage requirements, and has a high throughput cost, which is associated with ray tracing through different time slices. 
Although not presented here, several slow light simulations were generated to confirm that the fast light approximation does not seriously affect the library results (see e.g., \citealt{dexter_2010_SubmillimeterBumpSgr}; \citealt{moscibrodzka_2021_UnravelingCircularPolarimetric}).

\subsection{Radiative effects}
\label{sec:rad_pairs}

The pipeline we have described uses ideal GRMHD to generate the fluid simulations; we have assumed that M87 can be described by models in which radiative cooling is negligible so that it does not affect the dynamics of the plasma. This assumption was probed in \citetalias{EHTC_2019_5}, but it is likely that a full radiative treatment of the fluid simulation will be required in future analyses.

The M87 jet funnel may be populated by electron--positron pairs, produced either via cascades (e.g., \citealt{beskin_1992_FillingMagnetosphereSupermassive,levinson_2011_VariableTeVEmission,broderick_2015_HorizonscaleLeptonAcceleration,hirotani_2016_EnergeticGammaRadiation}) or drizzle \citep{moscibrodzka_2011_PAIRPRODUCTIONLOWLUMINOSITY,laurent_2018_ElectronPositronPair,kimura_2020_HadronicHighenergyEmission,wong_2021_PairDrizzleSubEddington,yao_2021_RadiationGRMHDSimulations}. Ideal GRMHD cannot produce unscreened electric fields and therefore cannot track pair cascades. Computing the cross sections for pair drizzle often requires a high-resolution sample of the radiation field, so it is expensive to track \emph{in situ} in GRMHD simulations and is often evaluated in a postprocessing step. Future study is warranted to investigate the signatures of pair plasma emission and whether or not pairs can populate the jet (see \citealt{broderick_2015_HorizonscaleLeptonAcceleration,anantua_2020_DeterminingCompositionRelativistic,emami_2021_PositronEffectsPolarized,yao_2021_RadiationGRMHDSimulations}).

\subsection{The accretion model}

The library presented in Section~\ref{sec:results} is not comprehensive. A more dense sampling of spin may enable a better understanding of how spin affects observables, particularly as spin approaches its maximal value $\left| \bhspin \right| = 1$. The transitory regime in which the magnetic flux increases from the comfortably SANE state toward the MAD state has been explored, but no dense parameter survey yet exists. The tilted-disk scenario merits further attention and study, even though it increases the dimensionality of the final parameter space by two.

A detailed study of the convergence properties of GRMHD simulations both with respect to spatial resolution and simulation duration would be valuable, as would a systematic survey of how the initial conditions affect the statistical properties of the fluid evolution and electromagnetic observables. 
 
\subsection{Viscosity}

GRMHD simulations typically assume a single ideal fluid and do not model the effects of fluid viscosity or resistivity.  While the magnetic Reynolds number in accretion flows is likely so high as to render resistivity irrelevant to the bulk dynamics of the flow, the very long mean free paths of accreting electrons and ions suggest it may be important to include the effect of viscosity, and thus heat conduction and pressure anisotropy. Note that explicit consideration of resistivity may still be important in studies of plasmoids and reconnection, which can contribute to the generation of flares and nonthermal particle injection (e.g., \citealt{ripperda_2021_BlackHoleFlares}).

Simulations treating pressure anisotropies and heat conduction find that, although the pressure anisotropies with respect to the magnetic field can be large in such systems, the overall structure of the accretion state remains similar to ideal simulations \citep{foucart_2017_HowImportantNonideal}. Thus, $230\,$GHz images of so-called extended GRMHD simulations are not expected to be drastically different from images of ideal GRMHD simulations. Nevertheless, including the effects of viscosity may alter the electron thermodynamics and particular dynamics of the flow.  Extended GRMHD simulations would have been prohibitively expensive to conduct at the range of parameters required for the current study and are the subject of current development \citep[see][]{chandra_2015_EXTENDEDMAGNETOHYDRODYNAMICSMODEL,chandra_2017_GrimFlexibleConservative,most_2021_DissipativeMagnetohydrodynamicsNonResistive,most_2021_ModelingGeneralrelativisticPlasmas}.

\subsection{Electron acceleration}

Single-fluid ideal GRMHD simulations assume that the plasma is well described by bulk variables that represent both the constituent electrons and ions; however, supermassive black hole accretion flows are likely collisionless, the electrons and ions are unequilibrated, and the dynamical imbalance between electron acceleration and cooling may produce a significantly nonthermal eDF. 

The influence of the nonthermal electrons on the observables can be roughly accounted for given a model for the eDF as a function of the local fluid parameters, since the radiative transfer coefficients associated with a given eDF can be computed and applied during postprocessing.
Rather than use a local post-hoc model for assigning the eDF, it is also possible to use a model for energy injection rates, e.g., motivated from PIC simulations (e.g., \citealt{sironi_2014_RelativisticReconnectionEfficient}), and track the nonthermal electron component directly in the fluid simulation by evolving the fraction of electrons associated with different energy bins in each simulation zone
\citep[e.g.,][]{ball_2016_ParticleAccelerationOrigin, chael_2017_EvolvingNonthermalElectrons,petersen_2020_NonthermalModelsInfrared}.

\section{Summary}
\label{sec:summary}

We have described \patoka, a numerical simulation pipeline that includes the \iharm, \ipole, and \igrmonty codes. \patoka can be used to generate spectra and fully polarimetric images of radiatively inefficient accretion flows around supermassive black holes. We have provided a brief sample of the \patoka data products that were used in the EHT analysis of total intensity and linear polarimetric data from the M87 black hole. Simulations and data produced from some or all of the \patoka pipeline have also been used in other works, including \citet{porth_2019_grmhdcomp,johnson_2020_UniversalInterferometricSignatures,lin_2020_FeatureExtractionSynthetic,palumbo_2020_DiscriminatingAccretionStates,gold_2020_VerificationRadiativeTransfer,wielgus_2020_wobble,ricarte_2020_DecomposingInternalFaraday,tiede_2020_VariationalImageFeature,gelles_2021_PolarizedImageEquatorial,wong_2021_JetDiskBoundaryLayer,ricarte_2021_BlackHoleMagnetic,ricarte_2022_spectralmaps}.

\acknowledgements

The authors thank Zachary Gelles, Michael Johnson, and Jordy Davelaar for many discussions surrounding the radiative transfer codes, and Christian Fromm and the anonymous referee for their careful reading of and helpful suggestions regarding the manuscript. This work was supported by National Science Foundation grants AST 17-16327, OISE 17-43747, AST 20-07936, AST 20-34306, by a Donald C.~and F.~Shirley Jones Fellowship to GNW, and by the US Department of Energy through Los Alamos National Laboratory. Los Alamos National Laboratory is operated by Triad National Security, LLC, for the National Nuclear Security Administration of the US Department of Energy (Contract No.~89233218CNA000001). MM acknowledges support by the NWO grant No.~OCENW.KLEIN.113. AR also acknowledges support from the Black Hole Initiative (BHI), enabled by grants from the Gordon and Betty Moore Foundation and the John Templeton Foundation. GNW also gratefully acknowledges support from the Institute for Advanced Study. This work has been assigned document release number LA-UR-21-28826.

\appendix

\section{Fishbone--Moncrief torus description}
\label{app:fmtorus}

\citet[][FM]{fishbone_1976_RelativisticFluidDisks} described a two-parameter solution of the relativistic Euler equations for an isentropic, stationary, axisymmetric, and purely azimuthal flow ($u^r = u^{\theta} = 0$) in a stationary, axisymmetric background. In Kerr, the Euler equations for the above scenario are
\begin{widetext}
\begin{align}
\ud \left[\ln h+\dfrac{1}{2}\ln\left(\frac{\Sigma\Delta}{A}\right)\right] = \dfrac{{u_{(\phi)}}^2}{2} \ud\left[\ln\left(\dfrac{\left(A \sin\theta\right)^2}{\Sigma^2\Delta}\right)\right] - 2 \bhspin u_{(\phi)}\left[1 + {u_{(\phi)}}^2\right]^{1/2}\frac{A\sin\theta}{\Sigma\sqrt{\Delta}} \, \ud \left[ \dfrac{r}{A}\right], 
\label{eqn:eulerEnthalpy}
\end{align}
\end{widetext}
where we write in Boyer--Lindquist (BL) coordinates, $h$ is the specific enthalpy of the fluid, $s$ is its specific entropy, $T$ is temperature, 
\begin{align}
    \Sigma &\equiv r^2 + \bhspin^2\cos\theta^2, \label{eqn:Sigmadefn} \\ 
    \Delta &\equiv r^2 - 2r + \bhspin^2, \label{eqn:Deltadefn} \\
    A &\equiv \left(r^2+\bhspin^2\right)^2 - \Delta\, \bhspin^2 \sin^2 \theta, \label{eqn:Adefn}
\end{align}
and the projections of $u^{\mu}$ onto the orthonormal vectors of the locally nonrotating frame are \citep{bardeen_1972_RotatingBlackHoles},
\begin{align}
u_{(\phi)} &= \sqrt{\Sigma/A} \, (1/\sin\theta) \, u_{\phi} \\
u_{(t)} &= -[1+{u_{(\phi)}}^2]^{1/2} \nonumber \\
&= \sqrt{A/\Sigma\Delta} \, u_{t} + 2 \bhspin r/(\Sigma\Delta A) \, u_{\phi} . \nonumber
\label{eqn:projFourvelocity}
\end{align}

\begin{figure*}[ht!]
\centering
\includegraphics[width=0.7\linewidth]{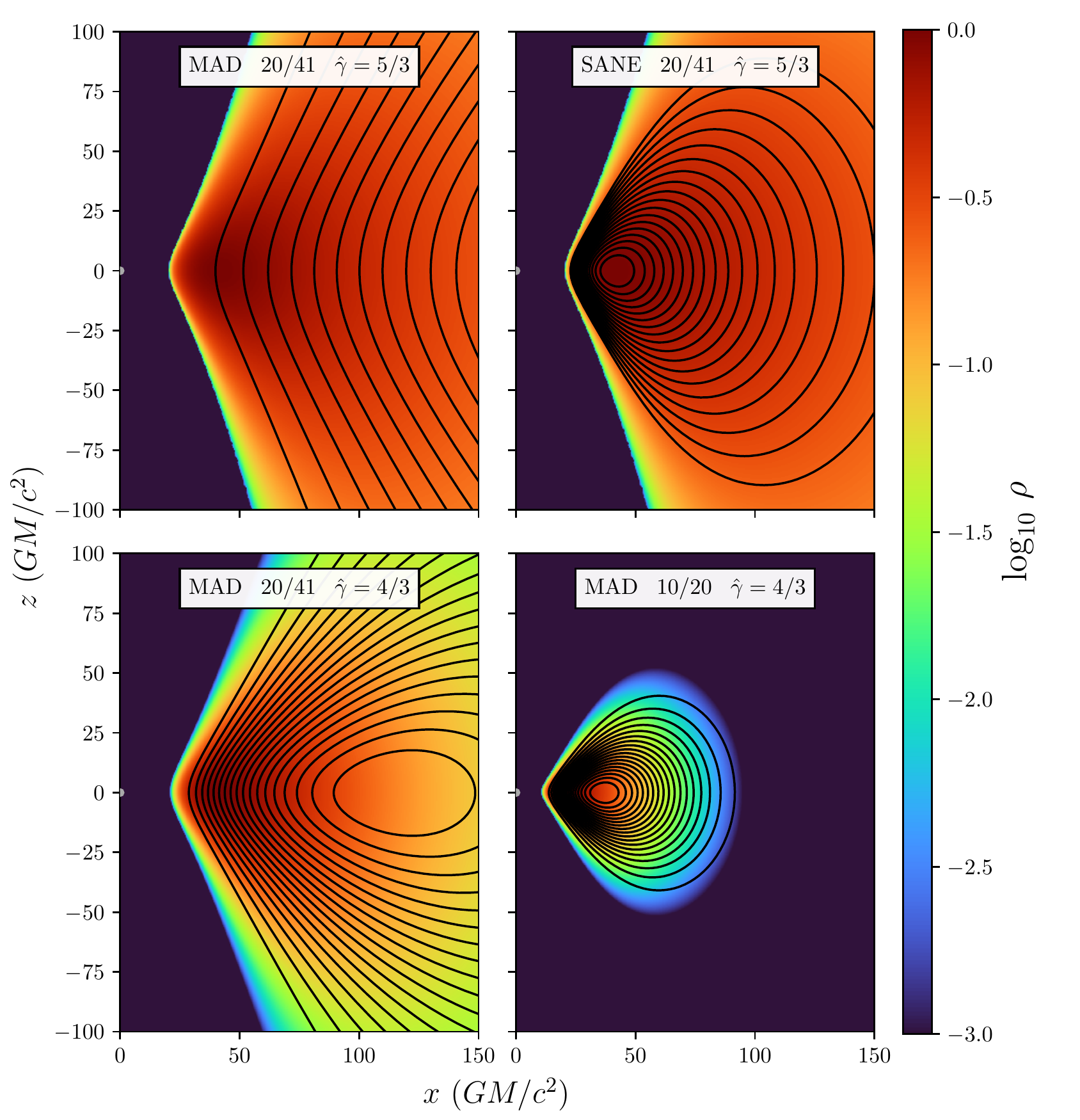}
\caption{Azimuthal slices of initial condition for fluid density (normalized and shown as logarithm) and magnetic field in MAD and SANE flows with different disk sizes (indicated by $r_{\rm in}$/$r_{\rm max}$) and adiabatic indices $\hat{\gamma}$ drawn from Table~\ref{table:grmhd_models}. The black hole corotates with the flow and has dimensionless spin parameter $\bhspin = 0.94$. The black hole event horizon is plotted as a filled gray circle at $x = z = 0$.}
\label{fig:harm_initial_conditions}
\end{figure*}

In our work, we take the solutions with constant $l = u_{\phi} u^t$. For these solutions, the fluid enthalpy can be expressed analytically as a function of $l$, $r$, $\bhspin$, $\theta$, and a parameter defining the location of the inner edge of the disk in the midplane $\rin$, 
\begin{widetext}
\begin{align}
\ln h = &\frac{1}{2} \ln \left[ \dfrac{1 + \left( 1 + 4\, l^2 \Sigma^2 \Delta / \left( A^2 \sin^2 \theta\right)\right)^{1/2}}{\Sigma \Delta / A} \right] - \dfrac{1}{2} \left( 1 + \dfrac{4 \,l^2 \Sigma^2 \Delta}{ A^2 \sin^2 \theta} \right)^{1/2} - \dfrac{2 \bhspin r l}{A} \\ \nonumber 
& - \dfrac{1}{2} \ln \left[ \dfrac{1 + \left( 1 + 4\, l^2  \Sigma_{\mathrm{in}}^2 \Delta_{\mathrm{in}}/ A_{\mathrm{in}}^2 \right)^{1/2}}{\Sigma_{\mathrm{in}}\Delta_{\mathrm{in}}/A_{\mathrm{in}}} \right] + \dfrac{1}{2} \left( 1 + \dfrac{4\, l^2 \Sigma_{\mathrm{in}}^2\Delta_{\mathrm{in}}}{A_{\mathrm{in}}^2}\right)^{1/2} + \dfrac{2 \bhspin \rin l}{A_{\mathrm{in}}^2}, \nonumber
\label{eqn:enthalpyequation}
\end{align}
\end{widetext}
where the second line is the negation of the first evaluated in the midplane at the inner edge of the torus, thus $\Sigma_{\mathrm{in}}$, $\Delta_{\mathrm{in}}$, and $A_{\mathrm{in}}$ are the expressions defined in Equations~\ref{eqn:Sigmadefn}--\ref{eqn:Adefn} with $r = \rin$.

\begin{figure*}
\centering
\includegraphics[width=\linewidth]{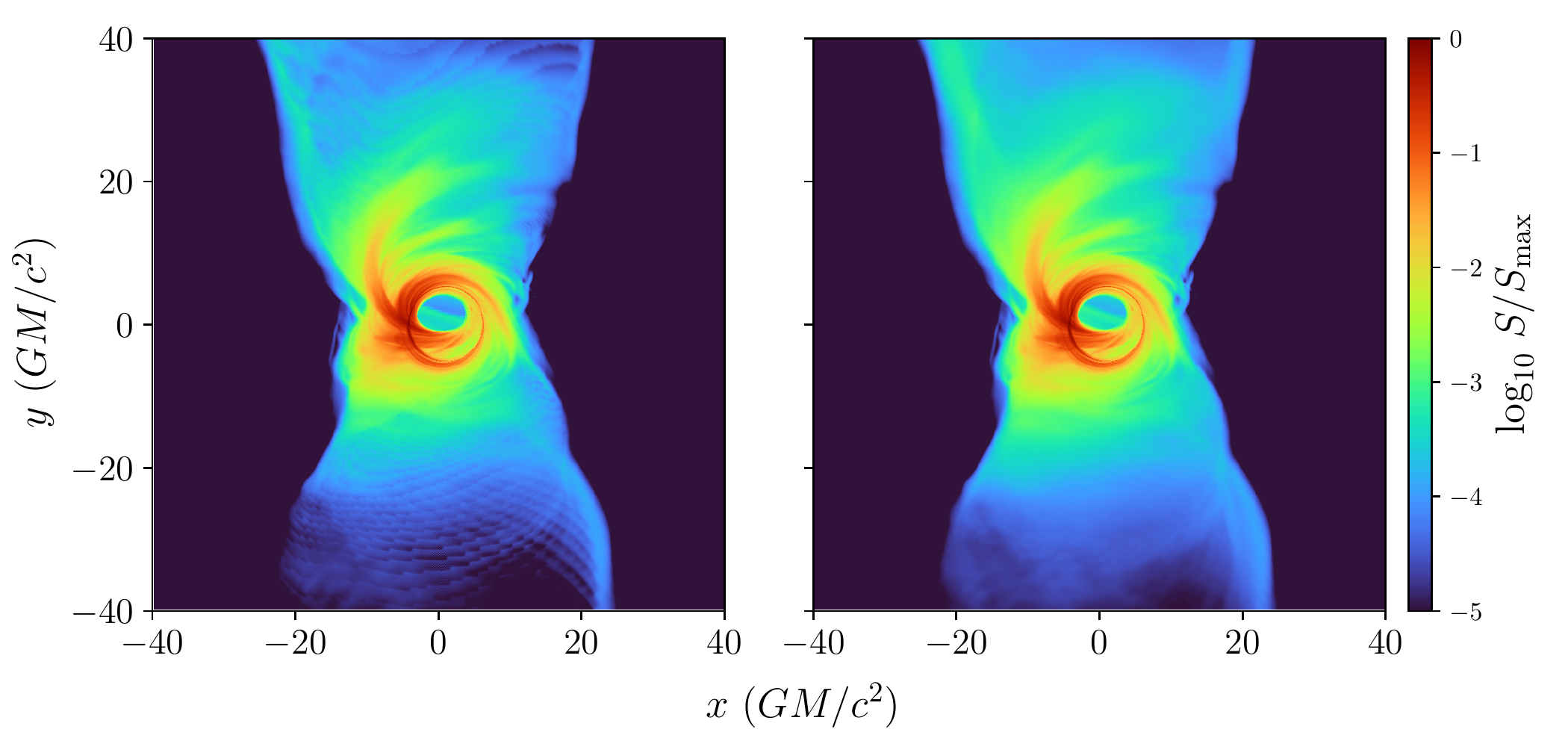}
\caption{Comparison of log-scaled total intensity image using zone-based vs.~interpolated-$\sigma$ cutoffs. In both panels, the cutoff is implemented by zeroing the electron number density. The sharp edges are from interpolation artifacts due to how linear interpolation deals with rapidly varying quantities on a nearly-regular grid. Notice that ridges are still present in the interpolated-$\sigma$ cutoff image; they are particularly visible near the bottom of the image. The ridge features may be accentuated when using an integrator that takes large steps.}
\label{fig:app:sigmacutoff_logscale}
\end{figure*}

In \iharm, we set the angular momentum density of the FM torus in terms of its value at $r_{\mathrm{max}}$, the radius of maximum pressure in the midplane
\begin{align}
l(r_{\mathrm{max}}) &= \dfrac{\bhspin^2-2\bhspin\sqrt{r}+r^2}{r^3\sqrt{2 \bhspin \sqrt{r}+r\left(r-3\right)}\left(\bhspin^2+r\left(r-2\right)\right)} \\
&\times\left[\dfrac{\left(-2\bhspin r\right)\left(\bhspin^2-2\bhspin\sqrt{r}+r^2\right)}{\sqrt{2 \bhspin \sqrt{r}+r\left(r-3\right)}} \right. \nonumber\\
&+ \left. \dfrac{\left(\bhspin+\sqrt{r}\left(r-2\right)\right)\left(r^3+\bhspin^2\left(r+2\right)\right)}{\sqrt{1+2 \bhspin r^{-3/2}-3/r}}\right] \Bigg|_{r=r_{\mathrm{max}}} \!\!\!\!\!\!\!\!\!\!\!\!\!\!\!. \nonumber
\end{align}

The region outside the torus is initialized to zero enthalpy. The remaining fluid variables are computed from the enthalpy using the ideal gas equation of state, the definition of specific enthalpy $h=1+\dfrac{\hat{\gamma}}{\hat{\gamma}-1}\dfrac{P}{\rho}$, and the isentropic condition, which enables us to equate the polytropic index and adiabatic index of the fluid. All grid zones that fall outside the torus are initialized to density and internal energy floor values.

For $u_{(\phi)}$, we use Equation 3.3 of FM
\begin{equation}
u_{(\phi)} = \sqrt{\frac{(1+4\,l^{2}\,\Sigma^{2}\Delta/(A \sin\theta)^2 )^{1/2}-1}{2}},
\label{eqn:u_phiExpression}
\end{equation}
so the azimuthal component of the four-velocity in BL coordinates is 
\begin{equation}
u^{\phi} = \frac{2\bhspin r}{\sqrt{\Sigma\Delta A}} \, \sqrt{1+{u_{(\phi)}}^2}+\sqrt{\dfrac{\Sigma}{A}}\dfrac{u_{(\phi)}}{\sin\theta} .
\label{eqn:uphiBL}
\end{equation}
The temporal component $u^t$ is computed using the four-velocity normalization condition $u^{\mu} u_{\mu}=-1$. 
Figure~\ref{fig:harm_initial_conditions} shows the fluid density and magnetic-field initial conditions for several example configurations.

\section{Primitive to conserved variables: conversion and inversion}
\label{app:harmprim}

Hyperbolic systems of conservation laws can be written in the form
\begin{align}
\partial_{t}U + \partial_{i}F^{i} = S,
\label{eqn:balanceLaw}
\end{align}
where $U$ is the conserved quantity, $F^{i}$ is the associated flux in the $i^{\mathrm{th}}$ spatial direction, and $S$ is the source term for the quantity. The partial differential equation is homogeneous when there are no source terms, leading to $U$ being conserved with time. Comparing Equation \ref{eqn:balanceLaw} with Equations (\ref{eqn:massConservation}--\ref{eqn:fluxConservation}), the vector of conserved variables for the governing equations of ideal GRMHD is
\begin{align}
\boldsymbol{U}\equiv\sqrt{-g} \,(\rho u^{t},\,{T^t}_{t},\,{T^t}_{i},\,B^{i}).
\end{align}

Computing the fluxes requires solving a local Riemann problem at zone faces. For higher-order reconstruction schemes, computing the fluxes is most convenient using an alternate vector of \emph{primitive} variables,
\begin{align}
\boldsymbol{P}\equiv(\rho,u,\tilde{u}^{i},B^{i}),
\end{align}
as the fluxes can be analytically computed from the primitives as $\boldsymbol{F}(\boldsymbol{P})$, unlike $\boldsymbol{F}(\boldsymbol{U})$. Although some primitives have the added benefit that it is easier to understand what they mean physically, note that $\tilde{u}^{i}$ has been chosen to improve numerical stability and is not the plasma three velocity:
\begin{align}
    \tilde{u}^i \equiv u^i + u^t g^{ti} \alpha^2,
\end{align}
where $\alpha^2 \equiv -1/g^{tt}$ is the lapse.

The conserved variables are complicated, analytical functions of the primitives. \iharm evaluates $\boldsymbol{U}(\boldsymbol{P})$ in two steps:
\begin{enumerate}
\item Calculate the fluid four-velocity $u^{\mu}$ and magnetic induction four-vector $b^{\mu}$ from $\boldsymbol{P}$,
\begin{align}
u^{t} &= \gamma/\alpha, \\
u^{i} &= \tilde{u}^{i} - \Gamma\beta^{i}/\alpha \label{eqn:primVelocity}, \\
b^{t} &= B^{i}u_{i}, \\
b^{i} &= (B^{i}+b^{t}u^{i})/u^{t},
\end{align}
where the Lorentz factor with respect to the normal observer can be computed as $\Gamma = \sqrt{1 + g_{ij} \tilde{u}_i \tilde{u}^j}$ and $\beta^i \equiv g^{ti} \alpha^2$ is the shift.
\item Calculate the stress--energy tensor (Equation \ref{eqn:mhdTensor}) from the primitives and four-vectors and use it to evaluate $\boldsymbol{U}$.
\end{enumerate}

The inverse operation for the primitives $\boldsymbol{P}(\boldsymbol{U})$ is performed using the ``$1D_{W}$'' scheme (\citealt{mignone_2007_eos_rmhd} and see also \citealt{noble_2006_PrimitiveVariableSolvers}). The matter conserved variables are nonlinear functions of the corresponding primitives, and there are no known analytic expressions for the inverse. The $1D_{W}$ scheme treats the conserved variables in the normal observer frame and defines a set of scalars, which can be constructed from $\boldsymbol{U}$, that reduces the 5D system to a 2D system. Of the two equations to be solved, one of them can be analytically inverted to obtain the Lorentz factor of the fluid. The other, which involves the energy density, is a nonlinear expression, and \iharm uses a 1D Newton-Raphson scheme to invert it. The magnetic-field primitives are equal to the corresponding conserved variables up to a factor of $\sqrt{-g}$.

\section{Interpolation and the \texorpdfstring{$\sigma$}{sigma} cutoff}
\label{app:sigmacutoff}

GRMHD codes are unable to robustly model fluid evolution in regions with high magnetization, so they often rely on limiting ``flooring'' procedures to ensure the numerical stability of the algorithm. In addition to modifying velocities, the flooring procedures introduce extra, artificial mass and energy, so they must be accounted for when performing ray tracing to generate electromagnetic observables. The standard procedure involves masking each fluid snapshot according to a threshold magnetization value $\sigma_{\mathrm{cut}}$. One would hope that the choice of $\sigma_{\mathrm{cut}}$ does not materially affect the simulated observables, and in many cases it does not; however, it has been observed to occasionally alter the morphology of images, e.g., in simulations of hot MAD flows (see \citealt{chael_2018_RoleElectronHeating}). 

\begin{figure}
\centering
\includegraphics[width=\linewidth]{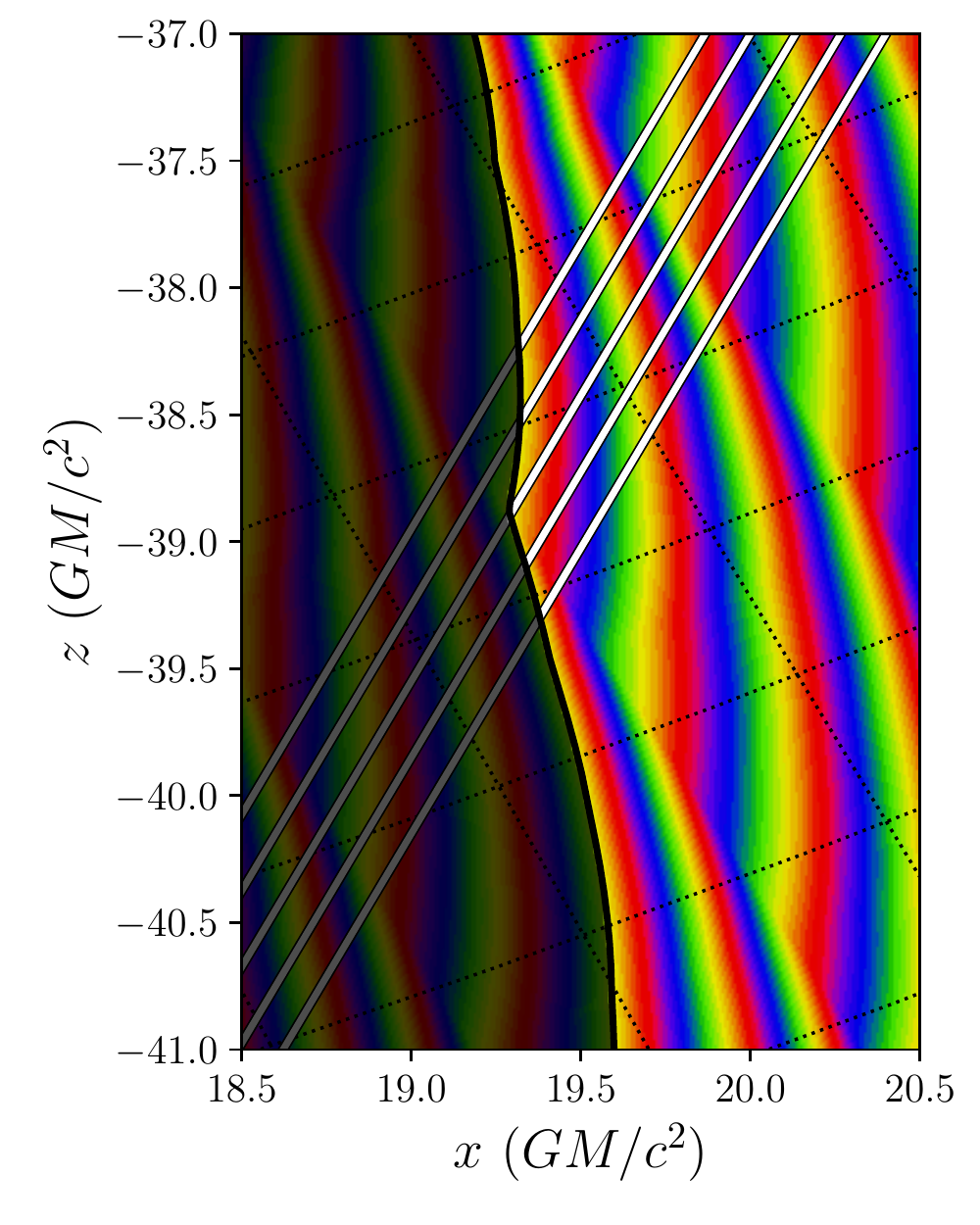}
\caption{Paths of five neighboring geodesics (white lines) plotted over bilinearly interpolated magnetization $\sigma$ (color, rapidly varying rainbow color scale). The $\sigma = 1$ surface is denoted with a solid black line, and zone boundaries are denoted by the dotted black lines. The magnetization decreases steadily from left to right but is defined at zone centers. The value of $\sigma$ at all other points is reconstructed using bilinear interpolation. The interpolation scheme produces a jagged transition at zone centers, and the slight deviations in path length produce ridges even when the value of $\sigma$ is determined from an interpolated value.}
\label{fig:app:sigmacut_trajectories}
\end{figure}

Depending on the interpolation algorithm used to reconstruct fluid variables at nonzone-centered locations and the step size of the integrator, the use of a $\sigma$ mask may also introduce ridging in simulated images that is due to the resolution of the underlying fluid snapshot. Figure~\ref{fig:app:sigmacutoff_logscale} shows severe (left) and minor (right) examples of the effect. The left panel shows the former default piecewise constant interpolation scheme for $\sigma$ that zeroed the zone-centered electron number density and introduces sharp boundaries in the image at low intensities. The right panel shows the same image but when $\sigma$ is (tri-)linearly interpolated and the mask is applied directly to the emissivity at each geodesic step. In general, the ridges produce artifacts in the Fourier domain that can completely change the signal at long baselines.

\begin{figure*}
\centering
\includegraphics[trim=20mm 0 15mm 0,width=0.9\linewidth]{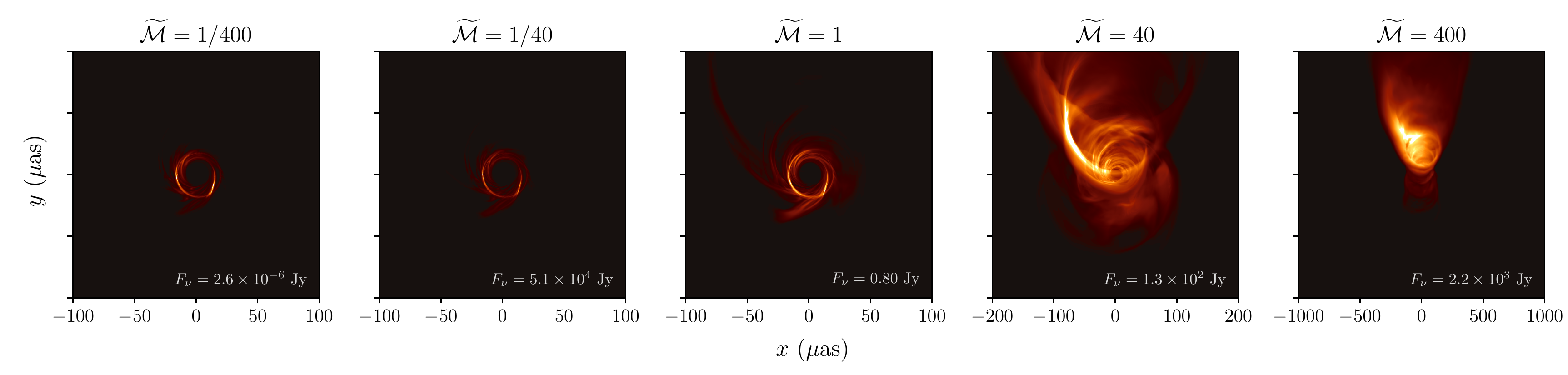}
\caption{Comparison of images produced from the same fluid snapshot across a range of mass density scales $\widetilde{\mathcal{M}} \propto \mathcal{M}$, where $\widetilde{\mathcal{M}} = 1$ is normalized to produce an M87-like flux density $F_{\nu} = 0.8\ \mathrm{Jy}$. Larger values of $\widetilde{\mathcal{M}}$ have higher optical depths; this effect becomes increasingly obvious at $\widetilde{\mathcal{M}} = 40$, where the flow begins to occlude the central flux depression. The high density scales chosen for the right two panels light up the funnel wall enough to substantially alter the image morphology, so the field of view for those two panels has been increased. Note that the models with large $\widetilde{\mathcal{M}}$ may not be physical since they are produced from fluid simulations that do not include all dynamically important effects at that mass accretion rate, e.g., cooling.}
\label{fig:grrt_massunit_images}
\end{figure*}

Even in the right panel, ridges can still be seen near the bottom of the image. This is due to the linear interpolation scheme, which produces values with discontinuous first derivative at zone centers. In multiple dimensions, the interpolation artifacts are particularly clear: Figure~\ref{fig:app:sigmacut_trajectories} shows the interpolated values of a smoothly varying scalar that has been sampled only at zone centers. When the interpolated value is used to mask the transfer coefficients along the geodesics (shown in white in the figure), then neighboring geodesics may have noticeably different path lengths.

\section{Varying the density scale \texorpdfstring{$\mathcal{M}$}{``M unit''}}
\label{app:lmd_degeneracy}

The output of the GRMHD simulation must be scaled to a particular choice of absolute, physical units in order to perform the radiative transfer calculation. The mass density scaling $\mathcal{M}$ is chosen to reproduce a target flux density, which is determined by observation. 
There is a mild degeneracy between the black hole mass $M$, the distance between the observer and the black hole $d_{\mathrm{src}}$, and the mass density unit $\mathcal{M}$. Although keeping $M/d_{\mathrm{src}}$ constant ensures that the angular size of the black hole on the sky remains fixed, changing $M$ alters the physical scale of the system. Thus at constant $\mathcal{M}$, increasing $M$ increases, e.g., the optical depth along a geodesic, which can change the image morphology. Thus, at large deviations of $M$ or $d_{\mathrm{src}}$, the degeneracy is broken, since $\mathcal{M}$ is used to fit the observational flux density constraint. A similar analysis as below but in the context of this degeneracy is presented in Appendix~A of \citet{roelofs_2021_BlackHoleParameter}, where both the black hole mass $M$ and the mass density unit $\mathcal{M}$ are rescaled at fixed flux density.

The image flux density $F_{\nu}$ is not a simple function of $\mathcal{M}$ primarily because of the complicated relationship between number density, magnetic-field strength, emission, and absorption: increasing $\mathcal{M}$ alters the magnetic-field strength, which nonlinearly changes the local emissivity; the local absorptivity along the line of sight increases as well, which can occlude parts of the flow. As a result, the relative brightness in different parts of the image changes.

Figure~\ref{fig:grrt_massunit_images} shows the images produced from a single MAD fluid snapshot with different values of $\widetilde{\mathcal{M}} \propto \mathcal{M}$, which are centered around the density scale that produces an M87-like flux density at $\widetilde{\mathcal{M}} = 1$. Different snapshots produce quantitatively distinct but qualitatively similar results. At low $\mathcal{M}$, the flow is optically thin and most emission is produced near the hole. As $\mathcal{M}$ is increased, optical depths through the disk and jet surface increase, and the primary observed emission shifts to the jet region. Note that, as $\mathcal{M}$ is increased, the accretion rate increases, cooling becomes more important, and the nonradiative assumption in the GRMHD is violated. Thus, the images with large $\mathcal{M}$ are likely not representative of physical accretion states.

Typically, $\mathcal{M}$ is determined automatically through a numerical root-finding procedure. Because the locus of emission shifts outward at high $\mathcal{M}$, the function $F_{\nu}(\mathcal{M})$ may appear to have multiple roots (potentially because the image is too small, the simulation domain is too small, or simply due to the changing structure of the emissivity and absorptivity as magnetic-field strengths increase). It is therefore important to verify that the resultant mass accretion rate is reasonable and that the synthetic observables are consistent with observation, e.g., by checking that the image is compact.

\begin{figure}
\centering
\includegraphics[trim=15mm 0 0 0,width=0.8\linewidth]{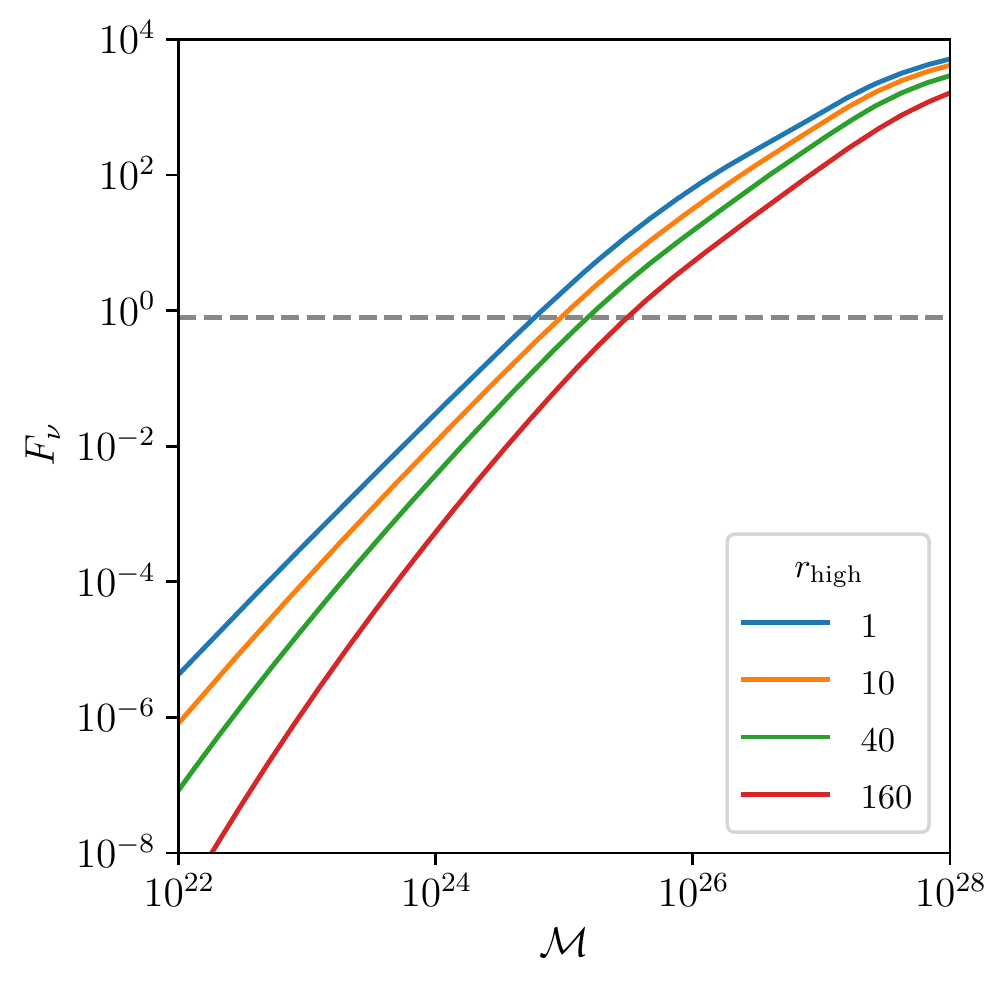}
\caption{Total flux density $F_\nu$ at $\nu = 230\,$GHz versus $\mathcal{M}$ for the same snapshot as Figure~\ref{fig:grrt_massunit_images} but for the four different values of $r_{\mathrm{high}}$. As before, note that the values of $F_\nu$ at large $\mathcal{M}$ may not be representative of physically meaningful scenarios, since cooling was not included in the fluid calculation.}
\label{fig:grrt_massunit_ftots}
\end{figure}

Figure~\ref{fig:grrt_massunit_ftots} shows the relationship between $F_{\nu}$ and $\mathcal{M}$ for the same snapshot shown in Figure~\ref{fig:grrt_massunit_images} for four different values of $r_{\mathrm{high}}$. Although the flux density typically increases with $\mathcal{M}$ and is relatively well-behaved, it is not possible to analytically determine which $\mathcal{M}$ corresponds to $F_{\nu} = 1$ in general. Nevertheless, given the correct value of $\mathcal{M}$ for a particular model, it is often possible to use that value as a seed estimate for the root-finding procedure when it is run on similar models.

How well do images at a given value $\mathcal{M}$ compare to nearby images? We quantify the difference between a trial image $K$ and the reference image $I$ according to three different metrics: the L1 ``taxicab'' metric, the mean-squared error across pixels, and the overall structural dissimilarity between the two images. In order to perform this comparison, we first rescale the trial image intensities so that both images have the same total flux density.

\begin{figure}
\centering
\includegraphics[trim={0mm 2mm 0mm 0mm},clip,width=0.95\linewidth]{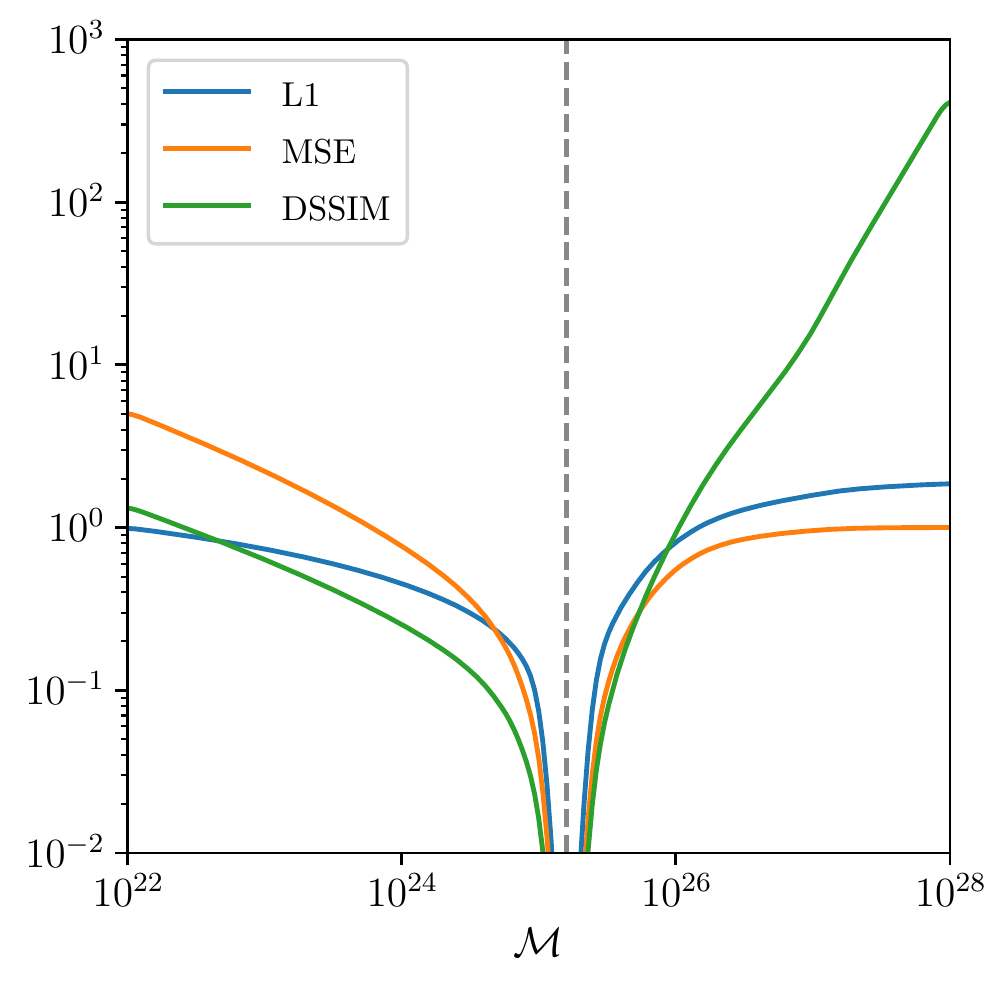}
\caption{Quantitative comparison between fiducial reference image with $\mathcal{M}$ chosen for $F_{\nu} = 0.8\ \mathrm{Jy}$ and images with other $\mathcal{M}$. All comparison images are drawn from a MAD $\bhspin=0.94$ model with $r_{\mathrm{high}}= 40$, as in Figure~\ref{fig:grrt_massunit_images}. As before, images produced with large $\mathcal{M}$ may not be representative of physically meaningful scenarios.}
\label{fig:grrt_massunit_msedssim}
\end{figure}

We compute the L1 error according to
\begin{align}
    L1(I,K) \equiv \dfrac{\sum \left| I_i - K_i \right|}{\sum I_i},
\end{align}
and we compute the mean-squared error as
\begin{align}
    \MSE(I,K) \equiv \dfrac{ \sum \left| I_i - K_i \right|^2}{ \sum {I_i}^2},
\end{align}
where the sums are taken over all pixels $i$ in the image. We compute the structural similarity index measure between the two images as
\begin{align}
    \SSIM(I, K) \equiv \dfrac{2\, \mu_I\, \mu_K}{\mu_I^2 + \mu_K^2} \dfrac{2\, \sigma_{IK}}{\sigma_I^2 + \sigma_K^2},
\end{align}
where
\begin{align}
    \mu_I &\equiv \sum \dfrac{I_i}{N} \\
    \sigma_I &\equiv \sum \dfrac{\left(I_i - \mu_I\right)^2}{N-1} \\
    \sigma_{IK} &\equiv \sum \dfrac{\left(I_i - \mu_I\right)\left(K_i - \mu_K\right)}{N-1}
\end{align}
are measures of the mean pixel intensity, variance, and covariance. We report $\SSIM$ in terms of the conventional structural dissimilarity
\begin{align}
    \DSSIM(I, K) = -1 + 1 / \left|\SSIM\right|  .
\end{align}
Figure~\ref{fig:grrt_massunit_msedssim} shows the L1, $\MSE$, and $\DSSIM$ measures over images from the $r_{\mathrm{high}} = 40$ model as compared to the M87-like reference image at $F_{\nu} = 0.8\ \mathrm{Jy}$.

\section{One-zone model comparison}
\label{app:polcomparison}

We have described the polarized and approximate ``unpolarized'' radiative transfer equations in Section~\ref{sec:radxfer}, and in Section~\ref{sec:codereview} we have outlined the observer-to-emitter and emitter-to-observer schemes used by \ipole and \igrmonty, respectively. We now use a simple one-zone model, like the one used in \citetalias{EHTC_2019_5}, to compare between the different schemes and codes.

Our model test problem comprises a ball in flat space with uniform electron number density $n_e = 2 \times 10^5\ \mathrm{cm}^{-3}$, temperature $\Theta_e = 10$, and magnetic field $B = 5\ \mathrm{G}$. The magnetic field has been oriented vertically, such that zero emission is produced along the two polar directions. The ball has radius $r_{\mathrm{out}} = 6.056 \times 10^{13}\ \mathrm{cm}$, corresponding to $100 \, GM/c^2$ for $M = 4.1 \times 10^6\, M_{\odot}$.

Both \igrmonty and \ipole are used to sample the full inclination-dependent synchrotron spectrum of the ball. \igrmonty only supports the approximate radiative transfer calculation, but we run \ipole in both modes for comparison. Since \ipole only produces images, we run it multiple times at different inclinations and frequencies to synthesize the full spectrum one point at a time. The $\nu L_{\nu}$ spectrum is computed from \ipole images by $\nu L_{\nu} = 10^{-23} \times 4 \pi \, d_{\mathrm{src}}^2 \, \nu F_{\nu}$. Here, $10^{-23}$ converts from the Jansky of spectral flux density, and $d_{\mathrm{src}} = 2.469 \times 10^{22}\ \mathrm{cm} = 8000\ \mathrm{kpc}$ is the distance between the camera and the center of the ball.

In \igrmonty, emission is sampled at zone centers, so it is important to adequately resolve the width of the photosphere to allow superphotons that undergo significant extinction to be treated correctly. In \ipole, images must both have high enough resolution to accurately reproduce gradients as well as a large enough field of view to see the full extent of the ball.

\begin{figure}
\centering
\includegraphics[trim=5mm 0 5mm 0, width=0.9\linewidth]{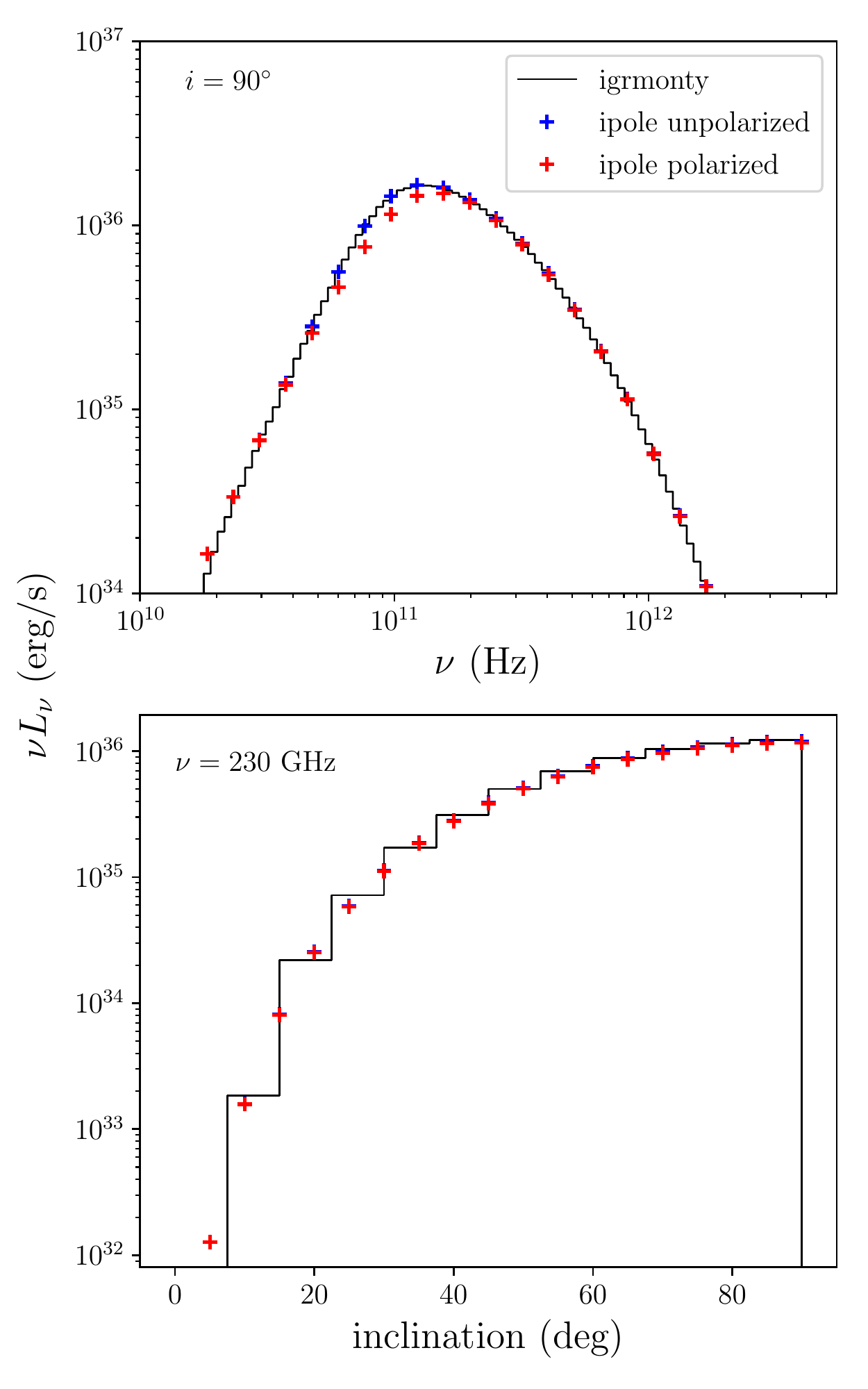}
\caption{Comparison of three radiative transport schemes for the one-zone model. Top panel shows flux densities at the edge-on $90^\circ$ inclination. Bottom panel shows flux densities for $\nu = 230\,$GHz. The approximate ``unpolarized'' transport scheme disagrees with the polarized one in the transition region as the ball becomes optically thick.}
\label{fig:grrt_pol_approx_diff}
\end{figure}

Figure~\ref{fig:grrt_pol_approx_diff} shows the results of the comparison. The \igrmonty and approximate ``unpolarized'' flux densities produced by \ipole agree across both inclination and frequency. The polarized and approximate solutions disagree near $\nu = 10^{11}\ \mathrm{Hz}$. The two methods disagree because the approximate solution does not account for the $\alpha_{Q,U,V}$ polarized absorptivities, and the effect is most significant in the transition between the regimes where the ball is optically thin ($\nu \gtrsim 10^{11}\ \mathrm{Hz}$) and optically thick ($\nu \lesssim 10^{11}\ \mathrm{Hz}$).

\section{Coordinates}
\label{app:fmkscoords}

GRMHD simulations are carried out on logically Cartesian grids, meaning that the internal coordinate representation forms a Cartesian grid, and non-Euclidean effects are accounted for through explicit treatment of geometric terms. Kerr--Schild (KS) coordinates $x^{\overline{\mu}} = \left(t, r, \theta, \phi\right)$ are chosen as the standard base coordinate system for fluid simulations because they are horizon penetrating. The exponential Kerr--Schild (eKS) coordinate system is one of the simplest extensions of KS coordinates and uses an exponential radial coordinate $x^1 \equiv \log(r)$, which increases the number of zones at small radii where both the relevant dynamical timescale is shorter and it is more important to recover the detailed dynamics of the flow. 

\begin{figure*}
\centering
\includegraphics[width=0.8374\linewidth]{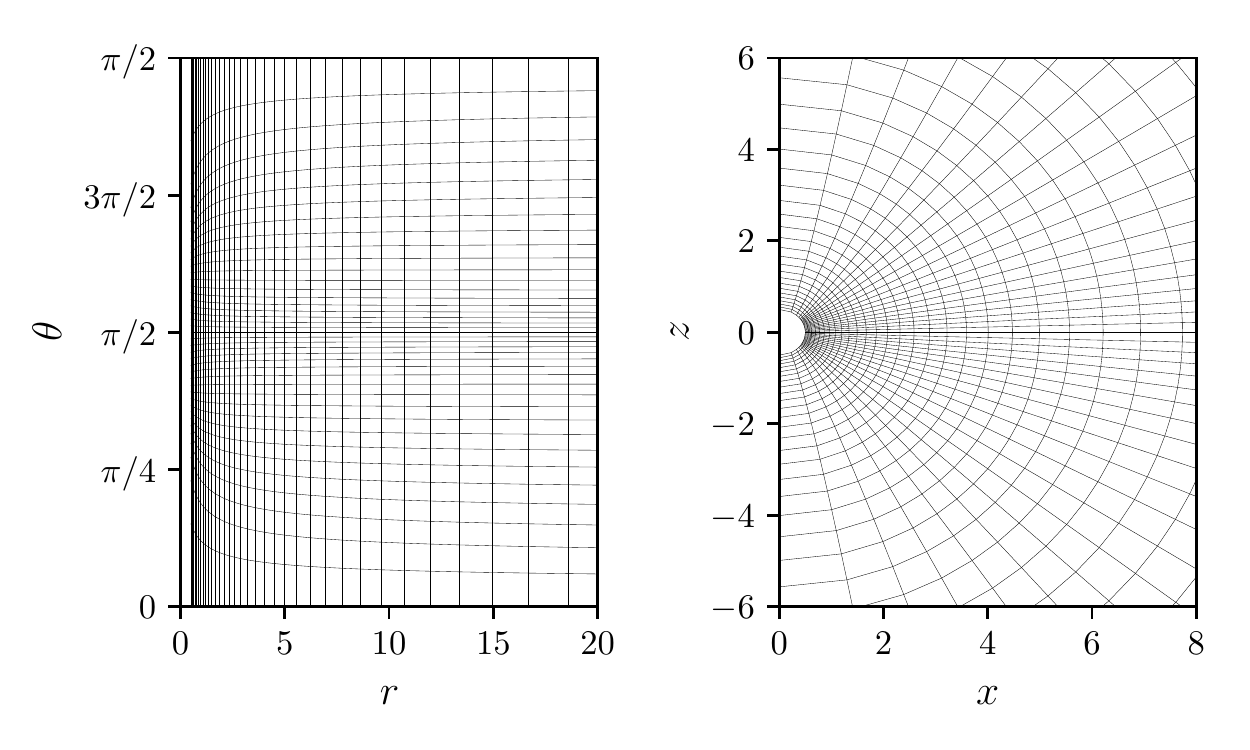}
\caption{Mapping between Kerr--Schild (KS) coordinates and funky modified Kerr--Schild (FMKS) coordinates. Left: lines of constant FMKS radial coordinate $x^1$ (vertical) and latitudinal coordinate $x^2$ (left-to-right) plotted versus KS radius $r$ and elevation $\theta$. Right: same as left but plotted in a Cartesian embedding with $x = r \sin \theta$ and $z = r \cos \theta$. FMKS coordinates concentrate resolution near the midplane $\theta = \pi/2$ and away from the poles $\theta = 0, \pi$ at small radii.}
\label{fig:coord_mapping}
\end{figure*}

The simulations in this paper were performed in funky modified Kerr--Schild (FMKS) coordinates $x^\mu = \left(x^0, x^1, x^2, x^3\right)$, which are an extension to the modified Kerr--Schild (MKS) coordinates introduced in \citet{gammie_2003_harm}, which are themselves an extension of eKS coordinates. Positive integer superscripts in this section should be interpreted as indices, not exponents. Coordinate modifications are generally chosen to both reduce computational cost and increase effective resolution by concentrating zones in regions of the domain where more interesting physics occurs---like the midplane and near the horizon at small radii---and through ``cylindrification,'' which expands ``unnecessary'' small zones---like those at the pole at small radii \citep[see][]{tchekhovskoy_2011_EfficientGenerationJets}. Each of FMKS, MKS, eKS, and KS is axisymmetric in $\phi$.

FMKS makes two modifications to the elevation coordinate $x^2$. The first reproduces MKS and increases the number of zones near the midplane by introducing a sinusoidally varying dependence of $\Delta (x^2)$ on $\theta$, as
\begin{align}
    \theta_g \equiv \pi x^2 +  \dfrac{1}{2} \left(1 - h\right)\sin\left(2 \pi x^2\right),
\end{align}
where $h$ is the midplane \emph{finification} parameter, which we set to $h = 0.3$. Note that there is no finification when $h=1$.

FMKS also introduces a cylindrification in $\theta$ whereby zones that are near the poles but at small radii have larger extent in $\theta$. This choice increases the length of the required numerical time step, which is set by the minimum of the signal-crossing time over all zones. The signal-crossing time in zones near the funnel often approaches the speed of light, and thus this fact combined with the structure of spherical geometry (which keeps the number of azimuthal zones constant regardless of $\theta$) results in many small zones with fast signal-crossing times. Thus, through cylindrification, we increase the size of the smallest zones and similarly gain an increase in time step. The cylindrification is achieved by defining 
\begin{align}
    \theta_j = N \left(2 x^2 - 1 \right) \left( 1 + \left(\dfrac{2 x^2 - 1}{B\left(1+\alpha\right)^{1/\alpha}}\right)^\alpha \right) + \pi / 2,
\end{align}
where $\alpha$ and $B$ are parameters and where
\begin{align}
    N = \dfrac{\pi}{2} \left( 1 + \dfrac{B^{-\alpha}}{1 + \alpha} \right)^{-1}
\end{align} 
is a normalization term. Finally, the KS co-latitudinal coordinate is
\begin{align}
    \theta = \theta_g + \exp\left[ -s \Delta x^1 \right] \left( \theta_j - \theta_g \right)
\end{align}
where $\Delta x^1 = x^1 - \log\left[r_{\mathrm{in}}\right]$ measures the FMKS distance from the inner edge of the simulation. In our simulations, we take $s = 0.5, B= 0.82,$ and $\alpha = 14$. Figure~\ref{fig:coord_mapping} shows an example of fluid zone boundaries versus KS coordinates for a grid with the above parameters.

We are not aware of an analytic inversion for $x^\mu(x^{\overline{\mu}})$, so a nonlinear root-finding step may be required to find the FMKS coordinates for a particular KS event, which is necessary when, e.g., setting the camera position during ray tracing.

\bibliography{main}
\bibliographystyle{aasjournal}

\end{document}